\documentclass[transaction]{IEEEtran}
\usepackage{graphicx}
\usepackage{epstopdf}
\usepackage[latin1]{inputenc}
\usepackage{amsmath,amssymb,amscd,latexsym,dsfont}
\usepackage[english]{babel}
\usepackage{tabularx,cite}
\usepackage{graphicx}
\usepackage{algpseudocode}
\usepackage{algorithm}
\usepackage{algorithmicx}
\usepackage{float}
\usepackage{multicol}
\usepackage{psfrag}
\usepackage{comment,psfrag,subfigure,enumerate}
\usepackage{color}
\usepackage{amsthm}
\usepackage{graphicx}
\usepackage{graphicx}
\usepackage{epstopdf}
\usepackage[latin1]{inputenc}
\usepackage{amsmath,amssymb,amscd,latexsym,dsfont}
\usepackage[english]{babel}
\usepackage{tabularx,cite}
\usepackage{graphicx}
\usepackage{algpseudocode}
\usepackage{algorithm}
\usepackage{algorithmicx}
\usepackage{float}
\usepackage{multicol}
\usepackage{mathtools}
\usepackage{psfrag}
\usepackage{comment,psfrag,subfigure,enumerate}
\usepackage{color}
\usepackage{amsthm}
\usepackage{multirow}
\ifCLASSINFOpdf
\else
\fi
\begin{document}
\IEEEoverridecommandlockouts
\IEEEpubid{\makebox[\columnwidth]{978-1-4799-5863-4/14/\$31.00 \copyright 2014 IEEE \hfill} \hspace{\columnsep}\makebox[\columnwidth]{ }}
\title{Fast HARQ over Finite Blocklength Codes: A Technique for Low-Latency Reliable Communication}

\author{\IEEEauthorblockN{Behrooz Makki, Tommy Svensson, \emph{Senior Member, IEEE}, Giuseppe Caire, \emph{Fellow, IEEE}, and Michele Zorzi, \emph{Fellow, IEEE} }
\thanks{Behrooz Makki, and Tommy Svensson are with Chalmers University of Technology, Gothenburg, Sweden, Email: \{behrooz.makki, tommy.svensson\}@chalmers.se. Giuseppe Caire is with the Department of Telecommunication Systems, Technical University of Berlin, Berlin 10623, Germany, and also with the Ming Hsieh Department of Electrical Engineering, University of Southern California, Los Angeles, CA 90089, USA, Email: caire@tu-berlin.de. Michele Zorzi is with the Department of Information Engineering, University of Padova, Padova, Italy, Email: zorzi@dei.unipd.it.}
\thanks{The work of Behrooz Makki and Tommy Svensson has received funding from the European Commission H2020 programme under grant agreement $n^{\circ}$671650 (5G PPP mmMAGIC project), and from the Swedish Governmental Agency for Innovation Systems (VINNOVA) within the VINN Excellence Center Chase.}
}
\maketitle
\vspace{-0mm}
\begin{abstract}
This paper studies the performance of delay-constrained hybrid automatic repeat request (HARQ) protocols. Particularly, we propose a fast HARQ protocol where, to increase the end-to-end throughput, some HARQ feedback signals and successive message decodings are omitted. {Considering {quasi-static channels and a bursty communication model}}, we derive closed-form expressions for the message decoding probabilities as well as the throughput, the expected delay and the {error} probability of the HARQ setups. The analysis is based on recent results on the achievable rates of finite-length codes and shows the effect of the codeword length on the system performance.  {Moreover, we evaluate the effect of various parameters such as imperfect channel estimation and hardware on the system performance.} As demonstrated, the proposed fast HARQ protocol reduces the packet transmission delay considerably, compared to state-of-the-art HARQ schemes. For example, with typical message decoding delay profiles and a maximum of $2,\ldots, 5$ transmission rounds, the proposed fast HARQ protocol can improve the expected delay, compared to standard HARQ, by {$27, 42, 52 $ and $60\%$}, respectively, independently of the code rate/fading model.
\end{abstract}



%
\IEEEpeerreviewmaketitle
\vspace{-0mm}
\section{Introduction}
The next generation of communication networks must provide high-rate reliable data streams for everyone everywhere at any time. Particularly, 5G targets peak data rates of the order of $10-100$ Gbps with an overall latency down to $1$ ms, e.g., \cite{backhaulexample}.
To meet such demands, it is required to use large bandwidths. In the radio frequency (RF) domain, 5G research is mainly concentrated on millimeter wave links with many transmit/receive antennas to obtain sufficiently large apertures/bandwidths/data rates. Moreover, the target is to maximize the overall throughput while at the same time  reducing the delays corresponding to, e.g., the signal processing, the channel state information (CSI) feedback process and the data transmission.

From another perspective, hybrid automatic repeat request (HARQ) is a well-known approach applied in wireless networks
to increase the data transmission reliability and efficiency \cite{6192276,7945856,MIMOARQkhodemun,throughputdef,5961851,a01661837,Tcomkhodemun}. HARQ systems can be viewed as channels with sequential feedback where, utilizing both forward error correction and error detection, the link reliability is improved by retransmitting the data that has experienced bad channel conditions. However, HARQ is commonly considered as a high-latency technique, with latency coming from multiple retransmissions, multiple message decoding at the receiver, and the acknowledgement/negative acknowledgement (ACK/NACK) signals {transmission}/processing delay. Therefore, to develop high-rate reliable communication over millimeter {wave} links, it is necessary to minimize the end-to-end transmission delay of the HARQ protocols. This is the motivation for our work, in which we design fast HARQ protocols with low end-to-end packet transmission delay.

The performance of HARQ protocols is studied in, e.g., \cite{MIMOARQkhodemun,throughputdef,5961851,a01661837,Tcomkhodemun}, where the throughput and the outage probability are investigated from {an} information theoretic point of view. Also, \cite{outageHARQ,greenkhodemun,4200959,5754756,6502167} propose different adaptive power and rate allocation schemes to improve the system performance.  Recently, we \cite{letterzorzikhodemun,7037211} and \cite{6831379} have studied the throughput and the power efficiency of HARQ protocols in the presence of short packets. Finally, e.g., \cite{7086845,948408,5734044} consider HARQ protocols in different millimeter wave communication setups where rate compatible codes are designed for incremental redundancy (INR) protocols with low encoding delay \cite{7086845}, a combination of  selective repeat (SR) and go-back-N (GBN) protocols is developed for $40$ GHz band radio channels \cite{948408}, and cross layer approaches based on adaptive modulation and HARQ are proposed for multiple-input-multiple-output (MIMO) Rician networks \cite{5734044}.


{The idea of using HARQ without waiting for the feedback on the basis of channel condition is not new and has been proposed in different works/standard specifications, e.g., \cite{6867300,phdthesisChaitanya,1327848}. Particularly, as the first approach to avoid excessive decoding effort in HARQ, \cite{1327848} proposes to avoid decoding in the first rounds using EXIT for low-density parity-check codes (LDPCs) INR as a proxy for the decoding process.} Also, in \cite{mmmagicdeliverableD41}, {we do pre-evaluations on the performance of millimeter {wave} systems avoiding unnecessary decoding in the rounds with low successful message decoding probability. Using   channel estimation to avoid HARQ-based feedback/decoding is  applied in} \cite{7556387} {as well. Here, the results are presented for the cloud radio access networks (C-RAN) where the typical base stations are divided into a remote radio head (RRH) that retains only radio functionalities and a baseband unit (BBU) that implements the rest of the protocol stack. In such systems, the fronthaul delay, i.e., the delay of transferring the data from the RRH to the BBU, signal processing at the BBU and sending feedback from the BBU to the RRH, increases the end-to-end delay of HARQ protocols considerably. Thus,} \cite{7556387} {introduces different schemes where, instead of message decoding by BBU, depending on the network architecture, with no need for message decoding, either the user or the RRH utilizes the channel estimate to predict if the BBU may or may not need retransmission. Here, the results are presented for Rayleigh {block}-fading channels where retransmission prediction is required in each round. {Reducing the HARQ processing delay in RAN has been studied in \cite{6867300} as well. Finally, see \cite{phdthesisChaitanya} for the review of different LTE-based schemes reducing the HARQ processing delay.}

In this paper, we concentrate on minimizing the transmission, the decoding and the feedback delay. Particularly, we develop a fast HARQ protocol where, to save on the end-to-end delay, 1) some ACK/NACK feedback signals are omitted and 2) the receiver attempts {to decode} the received messages not in all transmissions but only if it estimates {a} high successful decoding probability. In this way, we save on the decoding and the feedback delay. Also, we use
the recent results on the achievable rates of finite blocklength codes \cite{5452208,6802432,6572802} to analyze the system performance in the cases with short packets, which result in low transmission delay.

{With a limit on the maximum transmit power, the design problem
is cast in the form of minimizing the expected packet transmission delay subject
to an error probability constraint. Then, as a side result, we study the throughput of our proposed scheme as well.}
With {quasi-static channels, bursty communication} and {different} power amplifiers (PAs) efficiency models, we derive closed-form expressions for the successful message decoding probabilities, the expected delay and the {error} probability of the HARQ setups (Lemmas 1-4). Moreover,
we investigate the effect of the message decoding and the feedback process on the end-to-end packet transmission delay and compare the performance of the proposed scheme with that of the state-of-the-art HARQ protocols (Lemmas {5-6}). The developed conditions provide a benchmark for {the} usefulness of different decoding/feedback schemes in delay-constrained HARQ protocols.

The differences in the problem formulation, analysis and channel model make the problem solved in this paper completely different from the ones in the literature. Particularly, the paper is different from, e.g., \cite{6192276,7945856,MIMOARQkhodemun,6831379,6867300,phdthesisChaitanya,1327848,mmmagicdeliverableD41,throughputdef,5961851,a01661837,outageHARQ,Tcomkhodemun,greenkhodemun,4200959,5754756,6502167,letterzorzikhodemun,7037211,7086845,948408,5734044}, because we take the PAs properties, different fading models as well as the decoding and the feedback delays into account, and perform finite blocklength analysis of HARQ protocols. {As opposed to} \cite{7556387} {with Rayleigh {block}-fading channels and retransmission prediction in each round, we concentrate on quasi-static channels where the decision about the required number of retransmissions is made before multiple packet transmissions. Also, we present the results for both Rayleigh and Rician fading conditions with different levels of CSI at the receiver (CSIR) and PA models, and derive closed-form expressions for the error probability, throughput and delay which have not been presented before.}

The analytical and simulation results show that, compared to state-of-the-art HARQ protocols, the proposed fast HARQ protocol reduces the packet transmission delay considerably. As an example, consider {a} low/moderate signal-to-noise ratio (SNR) regime and {typical message decoding delay profiles}. Then, with a maximum of $2,\ldots, 5$ transmission rounds our proposed scheme improves the expected delay, compared to standard HARQ with message {decoding and feedback} in all rounds, by $27, 42, 52$ and $60\%$, respectively, independently of the code rate/fading model. Thus, the fast HARQ protocol can be effectively utilized in delay-constrained applications.  Then, the throughput and
the {error} probability are sensitive to the length of short signals while their sensitivity to the signals length decreases for long signals. Finally, the inefficiencies of the PAs affect the performance of HARQ protocols remarkably, and should be carefully considered in the network design.


\vspace{-0mm}
\section{System Model}
\begin{table*}
\caption{The definition of parameters.}\vspace{-0mm}
\begin{center}
\begin{tabular}{ |l|l|l|l|}
  \hline
  \small{Parameter} & \small{Definition}  &  \small{Parameter} & \small{Definition} \\
  \hline
  $g^i$ & \small{Channel gain {to the} $i$-th antenna }  & $P$ & \small{Transmission power}\\
  $N_\text{t}$ & \small{Number of transmit antennas}  & $G$ & \small{Sum channel gain}  \\
  $N_\text{r}$ & \small{Number of receive antennas} & $\Xi$ & \small{Relative expected delay gain} \\
  $M$& \small{Maximum number of HARQ transmissions} & $P^\text{max}$ & \small{Maximum output power}\\
  $(k,\Omega)$& \small{Fading parameters in Rician channels}& $P^\text{cons}$ & \small{Consumed power}\\
  $\Omega$& \small{Fading parameter in Rayleigh channels}& $q^i$ & \small{$i$-th quantization boundary}\\
  $\beta$& \small{Error probability constraint}& $L$ &\small{Sub-codewords length}\\
  $K$& \small{Nats per codeword}& $S^i$ &\small{$i$-th quantization region}\\
  $R$& \small{Initial transmission rate}& $\vartheta$ &\small{Power amplifier parameter}\\
  $\epsilon$& \small{Maximum efficiency of power amplifier}& $D$ &\small{Feedback delay}\\
  $\Lambda(\cdot)$& \small{Decoding delay {profile}}& $\bar\tau$ &\small{Expected delay}\\
  \hline
\end{tabular}
\end{center}\vspace{-0mm}
\end{table*}

%
{Table I summarizes the parameters used throughout the paper.} We study a point-to-point {single-input-multiple-output (SIMO)} setup with $N_\text{r}$ {receive} antennas (for discussions on MIMO setups, see Section III.A.1). {We present the results for bursty communications  model where there is an idle period between two successive packet transmissions and consider the stop-and-wait (SW) HARQ protocol motivated by its lowest complexity compared to, e.g., SR and GBN protocols.}

Let us define a packet as the transmission of a codeword along with all its possible HARQ-based retransmission rounds. We study quasi-static conditions where the channel coefficients remain constant during multiple packet transmissions and then change to other values based on their probability density functions (PDFs). This is {in harmony with, e.g.,} \cite{MIMOARQkhodemun,outageHARQ,greenkhodemun,a01661837,Tcomkhodemun}, {and} an acceptable assumption in the case of stationary or slow-moving users with bursty communication ({see Section III for discussions}). {Also, the assumption allows us to apply the finite blocklength results of} \cite{5452208,6802432,6572802} {in the performance evaluation of HARQ protocols.} In this way, the received signal is given by
\vspace{-0mm}
\begin{align}\label{eq:newchannelmodel}\vspace{-0mm}
{\textbf{Y}} = \textbf{H}{{X}} + \textbf{Z},
\end{align}
{where $\textbf{H}=[h^1,\ldots,h^{N_\text{r}}]\in \mathcal{C}^{1\times N_\text{r}}$ is the fading vector, ${X}\in\mathcal{C}^{1\times 1}$ is the transmitted signal and $\textbf{Z}\in \mathcal{C}^{1\times N_\text{r}}$ denotes the independent and identically distributed (IID) complex Gaussian noise vector}. Such a setup is of interest in, e.g., side-to-side communication between buildings/lamp posts, as well as in wireless backhaul point-to-point links where the trend is to introduce multiple antennas and thereby achieve multiple parallel streams, e.g., \cite{6515173}. Also, we define $g^i\doteq|h^i|^2,i=1,\ldots, N_\text{r},$ which are referred to as the channel gains in the following. {Our results are {initially} presented in the cases with perfect CSIR which is an acceptable assumption in quasi-static conditions. {Performance analysis in the cases with imperfect CSIR is studied in Section III.A.4.}

Let us represent the PDF and cumulative distribution function (CDF) of a random variable $X$ by $f_X(\cdot)$ and $F_X(\cdot)$, respectively. While the modeling of the millimeter wave-based links is well known for line-of-sight wireless backhaul links, it is still an ongoing research topic  for non-line-of-sight conditions \cite{5262304,7501500,6834753}. Particularly, different measurement setups have emphasized the near-line-of-sight propagation and the non-ideal hardware as two key challenges of such links. Here, we present analytical results for the Rayleigh and Rician channel models {as appropriate models for different use-case scenarios} \cite{mmwrician1,mmwrician2,mmwrician3,7481507,5433145}. {Particularly, } Rician fading {is a good model in} near-line-of-sight conditions and has been well established for different millimeter wave-based applications, e.g., \cite{mmwrician1,mmwrician2,mmwrician3,7481507,5433145}.
Rayleigh fading, on the other hand, is an appropriate model when there is no dominant line-of-sight propagation between the transmitter and the receiver, e.g., \cite{7481507,5433145}.
With a Rician model, the channel gain ${g^i,\forall i,}$ follows the PDF
\begin{align}\label{eq:eqRicianpdf}
f_{g^i}(x)=\frac{(k+1)e^{-k}}{\Omega}e^{-\frac{(k+1)x}{\Omega}}I_0\left(2\sqrt{\frac{k(k+1)x}{\Omega}} \right ),\forall i,
\end{align}
where $k$ and $\Omega$ denote the fading parameters and $I_n(\cdot)$ is the $n$-th order modified Bessel function of the first kind. Also, defining the sum channel gain $G=\sum_{i=1}^{N_\text{r}}{g^i},$ we have
\begin{align}\label{eq:eqRiciansumpdf}
f_{G}(x)&=\frac{(k+1)e^{-kN_\text{r}}}{\Omega}\left(\frac{(k+1)x}{kN_\text{r}\Omega}\right)^{\frac{N_\text{r}-1}{2}}\nonumber\\&\times e^{-\frac{(k+1)x}{\Omega}}I_{N_\text{r}-1}\left(2\sqrt{\frac{k(k+1)N_\text{r}x}{\Omega}} \right ).\end{align}

For Rayleigh fading conditions, on the other hand, we have $f_{g^i}(x)=\frac{1}{\Omega} e^{-\frac{1}{\Omega} x},\forall i,$ and
\begin{align}\label{eq:eqcdfRayleighsum}
f_G(x)=\frac{1}{\Omega^{N_\text{r}}(N_\text{r}-1)!}x^{N_\text{r}-1}e^{-\frac{1}{\Omega} x},
\end{align}
where $\Omega$ represents the fading parameter.


Finally, to take non-ideal hardware into account, we consider the state-of-the-art model for the PA efficiency where the output power at each antenna is determined according to
\cite{4160747}
\begin{align}\label{eq:ampmodeldaniel}
&\frac{P}{P^{\text{cons}}}=\epsilon\left(\frac{P}{P^{\text{max}}}\right)^\vartheta\Rightarrow  P=\sqrt[1-\vartheta]{\frac{\epsilon P^{\text{cons}}}{(P^{\text{max}})^\vartheta}}.
\end{align}
Here, $ P, P^{\text{max}}$ and $P^{\text{cons}}$ are the output, the maximum output and the consumed power, respectively, $\epsilon\in [0,1]$ denotes the maximum power efficiency achieved at $P=P^{\text{max}}$ and $\vartheta\in [0,1]$ is a parameter depending on the PA classes. Similar models as in (\ref{eq:ampmodeldaniel}) have been expressed by  \cite[Eq. (3)]{6725577}, and efficiency measurements of different classes of amplifiers have
indicated that the equation is indeed quite useful and accurate, e.g., \cite{4057776}. Therefore, in harmony with \cite{4160747,6725577,4057776}, we consider (\ref{eq:ampmodeldaniel}) as the PA model. Note that setting $\vartheta=0$ and  $\epsilon=1$  in (\ref{eq:ampmodeldaniel}) represents the cases with an ideal PA.
\vspace{-0mm}
\subsection{Data Transmission Model}
As the most promising HARQ approach leading to {the} highest throughput/lowest outage probability \cite{MIMOARQkhodemun,throughputdef,5961851,a01661837}, we consider INR HARQ with a maximum number $M-1$ {of} retransmissions, i.e., the data is transmitted {at most} $M$ times.
Using INR HARQ, $K$ information nats are encoded into a \emph{parent} codeword of length $ML$ channel uses. Then, the parent codeword is divided into $M$ sub-codewords of length $L$ channel uses which are sent in the successive transmission rounds (it is straightforward to extend the results to the cases with sub-codewords of different lengths). Thus, the equivalent data rate at the end of round $m$ is $R_{(m)}=\frac{K}{mL},R_{(0)}\doteq\infty,$ nats-per-channel-use (npcu). In each round, if the decoder is active, it combines all received sub-codewords to decode the message. The retransmission continues until the message is correctly decoded or the maximum permitted transmission round is reached.

The main idea of our proposed fast HARQ protocol is as follows (see Figs. 1-2). Receiving the signal in the first round, the receiver estimates the number of HARQ-based transmission rounds which are required to guarantee successful message decoding with high probability. Then, during these transmission rounds the receiver remains silent, i.e., it does not decode the message, while accumulating the received signals, and it sends no ACK/NACK feedback. The required number of transmissions is estimated by monitoring the sum channel gain $G$ defined in, e.g., (\ref{eq:eqRiciansumpdf}), or equivalently the received SNR $\gamma=PG$ with $P$ being the transmission power. In this way, with a maximum of $M$ transmission rounds, the range of the sum channel gain $G$ is quantized into $M$ regions as illustrated in Fig. 1. Let us denote the quantization boundaries by $q^i,i=0,\ldots,M,q^M=0,q^0=\infty,q^i\le q^{i-1},\forall i,$ and define the quantization regions $S^i=[q^i,q^{i-1} ),i=1,\ldots,M.$  At the end of round 1, if the receiver estimates the sum channel gain to be in region $S^m=[q^m,q^{m-1} ),$ it remains silent in rounds $i=1,\ldots,m-1,$ and accumulates the received signals. Then, in rounds $i=m,\ldots,M,$ the data transmission is performed in the standard HARQ-based fashion where the receiver decodes the received signal in each round, based on all signals received up to the end of that round, and sends ACK/NACK feedback signals depending on the message decoding status (Fig. 2). In this way, with the sum channel gain being in region $S^m$ the delays corresponding to message decoding in rounds $i=1,\ldots,m-1$ and $m-1$ feedback signals are saved by the proposed scheme.

{As seen in the following, the proposed scheme is applicable for different ranges of parameters and use-cases. More importantly, for every value of SNR/error probability, the proposed scheme can be considered as a bonus, because it improves the performance of the typical HARQ protocols with no extra cost at the transmitter/receiver. For these reasons, we do not limit our analytical and simulation results to a specific range of SNRs/error probabilities or use-case. However, as } a motivating example for the considered problem formulation, suppose an 802.11 type of communication \cite{6194400}, where when a user grabs the channel, it keeps the channel (including HARQ feedbacks) until it is released. In such systems, the user keeps sending more redundancy blocks till an ACK is fed back and the length of the initial transmission can be adapted to the channel conditions. Here, the focus is on short latency delivery of reliable packets, and the throughput is not the main metric of interest. For this reason, we concentrate on minimizing the error-limited packet transmission delay and, as a side result, study the throughput to complete the discussions. Also, considering the proposed scheme, it is interesting to note that:
\begin{itemize}
  \item There are practical HARQ protocols where {some kind of CSI estimation is used to save on the HARQ processing  delay } \cite{phdthesisChaitanya}. Here, we develop a theoretical framework for the analysis of HARQ protocols with reduced feedback and message decoding delays, and derive the ultimate performance gain of such systems.
  \item With our proposed scheme, at the end of the first round the receiver should inform the transmitter about the estimated number of required retransmissions (see Fig. 2). However, with quasi-static conditions, on which we concentrate, such quantized CSI feedback update is required after multiple packet transmissions and, consequently, its feedback overhead is negligible. For this reason, while it is straightforward to include, we ignore the cost of this quantized CSI feedback in our analysis. {As an alternative approach, our analysis is well applicable to the cases where, with no CSI, the transmitter keeps transmitting sub-codewords and listens to the possible ACK feedbacks. However, with the channel gain being in region $m$, the receiver starts decoding only in rounds $i\ge m$ and sends an ACK once the message is correctly decoded. }
  \item We consider the case where the pilot signals, required for channel estimation, are transmitted jointly with the information signal in the first transmission round. However, our derivations are also valid in the cases where the channel estimation signal is sent separately and the required number of transmission rounds is decided before data transmission.
  \item {We study {bursty} communications where there is an idle period between two successive packet transmissions (see} \cite{MIMOARQkhodemun} {for fundamental differences between the performance of HARQ protocols in bursty and continuous communication models). Also, we concentrate on SW HARQ scheme, motivated by its lowest complexity compared to SR and GBN protocols, and aim to reduce the end-to-end transmission delay subject to error probability constraints. Then, as seen in the following, the proposed scheme results in considerable delay reduction, because {it avoids message decoding if there is low probability for successful decoding}. Using, e.g., SR scheme in continuous communications scenarios, the white spaces between successive retransmissions are used for transmission of other packets. In such cases, our proposed scheme does not improve the end-to-end transmission delay. However, it still results in considerable reduction in the decoding implementation complexity and energy saving, as it reduces the probability of multiple unsuccessful message decodings.}
  \item {As in} \cite{7556387}, {the highest gain of our proposed scheme, compared to standard HARQ, is observed at low/moderate SNRs which is the range of interest in HARQ protocols with high probability for requiring retransmissions. At high SNRs, however, there is high probability that the message is  decoded in the first round, and the relative performance gain of such fast HARQ protocols decreases.}
\end{itemize}


In Section III, we study the performance of the proposed protocol.
Indeed, to find closed-form expressions for the {error} probability, the throughput and the expected delay, we need to implement approximation techniques. However, as shown in Section IV, the final conclusions are in harmony with the numerical simulations with high accuracy. {We} first review the results of \cite{5452208,6802432,6572802} on the achievable rates of finite blocklength codes as follows.

\subsection{On the Achievable Rates of Finite-length Codes \cite{5452208,6802432,6572802}}

Define an $(L,N,P,\delta)$ code as the collection of
\begin{itemize}
  \item An encoder $\Upsilon:\{1,\ldots,N\}\mapsto\mathcal{C}^L$ that maps the message $n\in\{1,\ldots,N\}$ into a length-$L$ codeword $x_n\in\{x_1,\ldots,x_N\}$ satisfying the power constraint
      \begin{align}\label{eq:code1}
\frac{1}{L}\left \| x_j \right \|^2\le P, \forall j.
\end{align}
  \item  A decoder $\Delta :\mathcal{C}^L\mapsto\{1,\ldots,N\}$ which satisfies the maximum error probability constraint
      \vspace{-0mm}
      \begin{align}\label{eq:code1}
\mathop {\max }\limits_{ \forall j}\Pr\left(\Delta (y(j))\ne j\right)\le \delta
\end{align}
with $y(j)$ denoting the channel output induced by the transmission of codeword $j$.
\end{itemize}
The maximum achievable rate of the code is given by
\begin{align}\label{eq:achievablerateeq1}
 R_\text{max}(L,P,\delta)=\sup\left\{\frac{\log N}{L}:\exists (L,N,P,\delta) \text{code}\right\} \,\,\text{(npcu)}.
\end{align}
Considering quasi-static conditions, \cite{5452208,6802432,6572802} have presented a very tight approximation for the maximum achievable rate (\ref{eq:achievablerateeq1})
which, for codes of rate $R$ npcu, leads to the error probability, e.g., \cite[Eq. (59)]{6802432}
\begin{align}\label{eq:errorfiniteblock2}
&\delta_\text{quasi-static}(L,R,P)\nonumber\\&\simeq E\left[Q\left(\frac{\sqrt{L}\left(\log(1+GP)+\frac{\log L}{2L}-R\right)}{\sqrt{1-\frac{1}{(1+GP)^2}}}\right)\right].
\end{align}
Here, $G$ is the instantaneous value of the channel gain and $E[\cdot]$ denotes the expectation with respect to the channel gain $G$.
Also, $Q(x)=\frac{1}{\sqrt{2\pi}}\int_x^\infty{e^{-\frac{t^2}{2}}\text{d}t}$ denotes the Gaussian $Q$-function. Since the approximation (\ref{eq:errorfiniteblock2}) has been shown to be very tight for moderate/large values of $L$ \cite{5452208,6802432}, for simplicity we will assume that they are exact in the following.
%
{Also, because for a broad range of parameter settings {and SNRs} the third-order approximation term $\frac{\log L}{2L}$ in} (\ref{eq:errorfiniteblock2}) {is much smaller than the {other terms in the numerator of the $Q$-function} and to simplify the expressions, we ignore it in our analysis. However, all results can be rewritten for the cases considering $\frac{\log L}{2L}$ {(see Fig. 3)}.}





\begin{figure}
\centering
  \includegraphics[width=0.95\columnwidth]{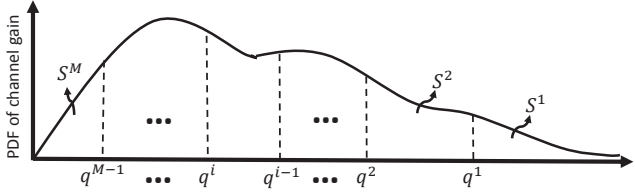}\\\vspace{-2mm}
\caption{An example of quantization boundaries. If the channel gain falls into the $i$-th quantization region, i.e., $G\in S^i=[q^{i}, q^{i-1})$, the receiver remains silent in rounds $1,\ldots, i-1$, and performs standard HARQ in rounds $i,\ldots, M.$}\vspace{-2mm}\label{figure111}
\end{figure}
\begin{figure*}
\centering
  \includegraphics[width=1.95\columnwidth]{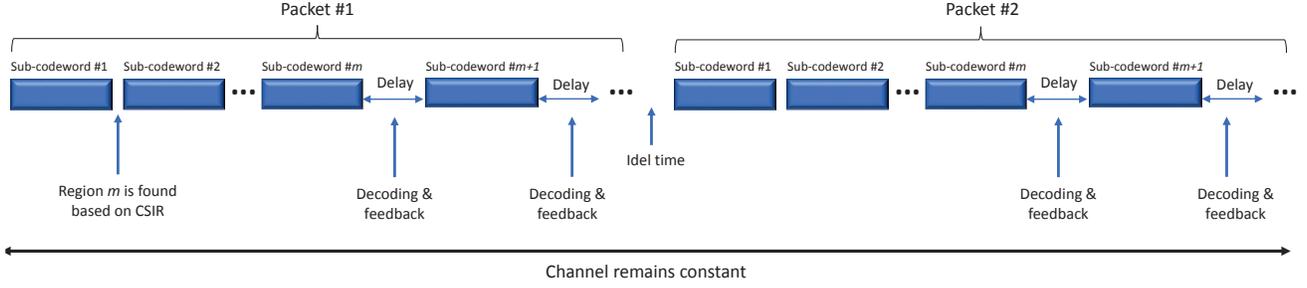}\\\vspace{-1mm}
\caption{Schematic of the packet transmission in the proposed approach.}\vspace{-1mm}\label{figure111}
\end{figure*}


\vspace{-0mm}
\section{Analytical results}
Let $D$ (in channel uses) denote the HARQ feedback delay in each round and $\Lambda(L)$ represent the delay for decoding a message of length $L$ channel uses. {In practice, $\Lambda(\cdot)$ depends on the coding scheme, number of iterations in the iterative decoder, etc. For this reason, we do not specify the characteristics of $\Lambda(\cdot)$.} Suppose that the instantaneous channel gain falls into the $m$-th quantization region, i.e., $G\in S^m$. If the data transmission is stopped at the end of round $i\ge m$, the total number of channel uses is
\begin{align}\label{eq:eqconditionalexpecteddelay}
\tau (i|G\in S^m)=\left\{\begin{matrix}
 iL+\Lambda (iL)+(i-m+1)D & \text{if } i\ne M, \\
 ML+\Lambda(ML)+(M-m)D & \text{if } i=M.
\end{matrix}\right.
\end{align}
This is based on the fact that 1) with sum channel gain being in the $m$-th quantization region the HARQ-based message decoding starts after $m$ transmissions and 2) in each transmission round $i\ge m$, except the last {one}, an ACK/NACK signal is fed back to the transmitter. In this way, with some manipulations, the expected number of channel uses in each packet transmission period given that the channel gain is in the $m$-th quantization region is found as
\begin{align}\label{eq:eqconditionalexpecteddelay2}
E\{\tau|&G\in S^m\}\nonumber\\&=\left\{\begin{matrix}
mL+\Lambda(mL)+\sum_{i=m+1}^M{\left(L+\Lambda(iL) \right )\phi_{i-1}^m}\,\,\,\,\,\,\,\,\,\,\,\,\,\,\,\,\\+D\left(1+\sum_{i=m+1}^{M-1}{\phi_{i-1}^m} \right ) \,\,\,\,\,\,\,\,\,\,\,\,\,\,\,\,\, \text{if } m<M,\,\,\,\,\,\,\,\,\,\,\,\,\,\,\,\,\,\,\,\,\,\,\,\,\,\, \\
ML+\Lambda(ML) \,\,\,\,\,\,\,\,\,\,\,\,\,\,\,\,\,\,\,\,\,\,\,\,\,\,\,\,\,\,\,\,\,\,\,\,\, \text{if } m=M,\,\,\,\,\,\,\,\,\,\,\,\,\,\,\,\,\,
\end{matrix}\right.
\end{align}
where $\phi_i^m$ denotes the probability that the message is not correctly decoded up to the end of the $i$-th round given that the channel gain is in the $m$-th quantization region. Moreover, the total expected packet transmission delay is obtained as
\begin{align}\label{eq:eqexpecteddelay}
\bar \tau&=\sum_{m=1}^M{\Pr\left(G\in S^m \right )E\{\tau|G\in S^m\}}
\nonumber\\&=\sum_{m=1}^M{\Pr\left(G\in S^m \right )\left(mL+\Lambda(mL) \right )}\nonumber\\&+\sum_{m=1}^{M-1}{\Pr\left(G\in S^m \right )\left(\sum_{i=m+1}^M{\left(L+\Lambda(iL) \right )\phi_{i-1}^m} \right )}
\nonumber\\&+D\sum_{m=1}^{M-1}{\Pr\left(G\in S^m \right )\left(1+\sum_{i=m+1}^{M-1}{\phi_{i-1}^m}\right)}
\nonumber\\&
=\sum_{m=1}^M{\Pr\left(G\in S^m \right )\left(mL+\Lambda(mL) \right )}\nonumber\\&+\sum_{m=1}^{M-1}{\sum_{i=m+1}^M{\left(L+\Lambda(iL) \right )\theta_{i-1}^m} }\nonumber\\&
+D\sum_{m=1}^{M-1}{\Pr\left(G\in S^m \right )}
+D\sum_{m=1}^{M-1}{\sum_{i=m+1}^{M-1}{\theta_{i-1}^m}}
\end{align}
Here, $\theta_i^m$ is the probability that the channel gain is in the $m$-th quantization region and the message is not correctly decoded up to the end of the $i$-th round. Also, $\Pr\left(G\in S^m \right )$ denotes the probability that the channel gain falls into the $m$-th quantization region. Thus, to derive the expected delay, we need to calculate the probabilities $\Pr\left(G\in S^m \right )$ and $\theta_i^m,\forall i,m.$ Using the CDF of the sum channel gain $G$,  we have
\begin{align}\label{eq:eqprobSm}
\Pr\left(G\in S^m \right )=\int_{q^m}^{q^{m-1}}{f_G(x)\text{d}x}=F_G\left(q^{m-1}\right)-F_G\left(q^m\right).
\end{align}
Moreover, considering INR HARQ and the properties of the power amplifier (\ref{eq:ampmodeldaniel}), we can use (\ref{eq:errorfiniteblock2}) to obtain $\theta_i^m$ by
\begin{align}\label{eq:eqthetaim}
\theta_i^m=\int_{q^m}^{q^{m-1}}{f_G(x)Q\Bigg(\frac{\sqrt{iL}\left(\log\left(1+x\sqrt[1-\vartheta]{\frac{\epsilon P^{\text{cons}}}{(P^{\text{max}})^\vartheta}}\right)-\frac{K}{iL}\right)}{\sqrt{1-\frac{1}{\left(1+x\sqrt[1-\vartheta]{\frac{\epsilon P^{\text{cons}}}{(P^{\text{max}})^\vartheta}}\right)^2}}}\Bigg)\text{d}x},
\end{align}
which is based on the fact that, given {that} the channel gain falls into the $m$-th quantization region, the sum channel gain PDF is given by
\vspace{-0mm}
\begin{align}\label{eq:eqgivenchannelgain}
f_G(x|G\in S^m)=\left\{\begin{matrix}
\frac{1}{F_G(q^{m-1})-F_G(q^m)}f_G(x), &\text{if } x\in S^m\\
0 &\text{otherwise.}
\end{matrix}\right.
\end{align}
{Also,} (\ref{eq:eqthetaim}) {is based on the upper bound $\Pr(A_m, A_{m+1},\ldots, A_i|G\in S^m)\le\Pr(A_i|G\in S^m), i\ge m$,  with $A_m$ being the event of error in the $m$-th transmission which, as shown in} \cite{7556387,approxerrorfinite}, {is very tight for moderate/large values of $L$ on which we concentrate. Then,} it is straightforward to show that the proposed fast HARQ protocol leads to the same {error} probability as in standard HARQ, i.e.,
\begin{align}\label{eq:eqoutageprob1}
&\Pr\left(\text{error}\right)=\sum_{m=1}^M{\Pr\left(G\in S^m\right)\Pr\left(\text{error}|G\in S^m\right)}
\nonumber\\&=
\sum_{m=1}^M{\int_{q^m}^{q^{m-1}}f_G(x)\times} \nonumber\\&\,\,\,\,\,\,\,\,\,\,\,\,\,\,\,\,\,\,\,\,\,\,\,\,\,\,\,\,\,{Q\Bigg(\frac{\sqrt{ML}\left(\log\left(1+x\sqrt[1-\vartheta]{\frac{\epsilon P^{\text{cons}}}{(P^{\text{max}})^\vartheta}}\right)-\frac{K}{ML}\right)}{\sqrt{1-\frac{1}{\left(1+x\sqrt[1-\vartheta]{\frac{\epsilon P^{\text{cons}}}{(P^{\text{max}})^\vartheta}}\right)^2}}}\Bigg)\text{d}x}
\nonumber\\&=
{\int_{0}^{\infty}f_G(x)Q\Bigg(\frac{\sqrt{ML}\left(\log\left(1+x\sqrt[1-\vartheta]{\frac{\epsilon P^{\text{cons}}}{(P^{\text{max}})^\vartheta}}\right)-\frac{K}{ML}\right)}{\sqrt{1-\frac{1}{\left(1+x\sqrt[1-\vartheta]{\frac{\epsilon P^{\text{cons}}}{(P^{\text{max}})^\vartheta}}\right)^2}}}\Bigg)\text{d}x}.
\end{align}
Finally, note that, as demonstrated in, e.g., \cite{MIMOARQkhodemun,throughputdef,5961851,a01661837},  for different channel
models, the throughput of different HARQ protocols can be written as
\begin{align}\label{eq:eqthroughputdef}
\eta=\frac{K\left(1-\Pr\left(\text{error} \right )\right)}{\bar\tau}.
\end{align}
In this way, the expected packet transmission delay (\ref{eq:eqexpecteddelay}), the {error} probability (\ref{eq:eqoutageprob1}) and the throughput (\ref{eq:eqthroughputdef}) of HARQ protocols are monotonic functions of the probabilities
\begin{align}\label{eq:eqprobZ}
&\mathcal{Y}(a,b,n,L)\nonumber\\&=\int_a^b{f_G(x)Q\Bigg(\frac{\sqrt{nL}\left(\log\left(1+x\sqrt[1-\vartheta]{\frac{\epsilon P^{\text{cons}}}{(P^{\text{max}})^\vartheta}}\right)-\frac{K}{nL}\right)}{\sqrt{1-\frac{1}{\left(1+x\sqrt[1-\vartheta]{\frac{\epsilon P^{\text{cons}}}{(P^{\text{max}})^\vartheta}}\right)^2}}}\Bigg)\text{d}x},
\end{align}
and, to analyze the system performance, the final step is to derive (\ref{eq:eqprobZ}). For different fading conditions, (\ref{eq:eqprobZ}) does not have a closed-form expression. For this reason, Lemmas 1-4 are used to approximate the probabilities $\mathcal{Y}(a,b,n,L),\forall a,b,n,L,$ as follows.

\textbf{\emph{Lemma 1:}} Consider {a} Rician fading model. For moderate/large number of antennas, which is of interest in millimeter wave communication,  the sum gain $G$ is approximated by {a} Gaussian random variable $\mathcal{Z}\sim\mathcal{N}(N_\text{r}\zeta,N_\text{r}\nu^2)$ with $\zeta=\mathcal{S}(1)$, $\nu^2=\mathcal{S}(2)-\mathcal{S}(1)^2$ and $\mathcal{S}(n)\doteq\left(\frac{\Omega}{k+1}\right)^n\Gamma\left(1+n\right)\prescript{}{1}F_{1}(n+1;1;k)$. 
Here, $\prescript{}{s}F_{t}(\cdot)$ denotes the generalized hypergeometric function and $k,\Omega$ are the fading parameters as defined in (\ref{eq:eqRicianpdf}).

\begin{proof}
{See Appendix A.}
\end{proof}

{\textbf{\emph{Lemma 2:}} With {a} Rician fading model, the sum gain $G$ is approximated by {a} Gamma random variable $ \mathcal{B}(x)=\frac{s_0^{s_1}}{\Gamma(s_1)}x^{s_1-1}e^{-s_0x}$ with $s_0=\frac{\zeta}{\nu^2}$, $s_1=\frac{N_\text{r}^2\zeta}{\nu^2}$ and $\zeta$ and $\nu$ given in Lemma 1. {Also, the same point holds for the cases with Rayleigh fading where the parameters of the Gamma random variable are given by $s_0=\frac{1}{\Omega}$ and $s_1=N_\text{r}.$}}
\begin{proof}
{The proof follows the same procedure as in Lemma 1 and the fact that, using the CLT for causal functions, the sum of positive independent random variables can be approximated by a Gamma distribution with parameters given in} \cite[Section 4.1]{GammaCLT}.
\end{proof}

{Lemmas 1-2 lead} to the following corollary statements:
\begin{itemize}
  \item With {a} moderate/large number of antennas, {the same {error} probability, expected delay and throughput are achieved in the cases with different fading models as long as the long-term channel parameters are properly set. This is because for different fading models one can use the same approach as in {Lemmas 1-2} and replace the channel gain by Gaussian {or Gamma} variables whose means and variances depend on the long-term channel characteristics}.
  \item Using {the} same approach as in {Lemmas 1-2}, we can map the {SIMO}-HARQ system into an equivalent single-input-single-output (SISO)-HARQ setup whose fading follows {Gaussian or Gamma PDFs with parameters given in Lemmas 1-2, respectively}. This is interesting because the mapping simplifies the performance analysis of {SIMO}-HARQ and makes it possible to extend many theoretical results of SISO-HARQ setups to {SIMO}-HARQ.
\end{itemize}

{For simplicity, Lemma 3 uses the CLT-based approximation approach of Lemma 1 to find} (\ref{eq:eqprobZ}) {for the cases with Rician channels. However, the same approach as in Lemmas 3 can be applied to approximate} (\ref{eq:eqprobZ}) {based on the Gamma-based approximation approach of Lemma 2.}

\textbf{\emph{Lemma 3:}} For Rician channel models, the probabilities $\mathcal{Y}(a,b,n,L),\forall a,b,n,L,$ are approximated by {$\mathcal{Y}(a,b,n,L)\simeq \mathcal{U}(a,b,n,L)$ where}
\begin{align}\label{eq:eqprobZapprox}
&\mathcal{U}(a,b,n,L)
\nonumber\\&=Q\left(\frac{\min\left(a,c_n\right)-N_\text{r}\zeta}{N_\text{r}\nu^2}\right)-Q\left(\frac{{\min\left(b,c_n\right)}-N_\text{r}\zeta}{N_\text{r}\nu^2}\right)
\nonumber\\&
+\frac{\frac{1}{2}+\alpha_n\mu_n-\mu_n N_\text{r}\zeta}{2}\times \nonumber\\&\left(1-2Q\left(\frac{{\max\left(\min\left(b,d_n\right),c_n\right)}-N_\text{r}\zeta}{\sqrt{N_\text{r}\nu^2}}\right)\right)
\nonumber\\&+\frac{\mu\sqrt{2N_\text{r}\nu^2}}{\sqrt{\pi}}e^{-\frac{\left({\max\left(\min\left(b,d_n\right),c_n\right)}-N_\text{r}\zeta\right)^2}{2N_\text{r}\nu^2}}
\nonumber\\&
-\frac{\frac{1}{2}+\alpha_n\mu_n-\mu_n N_\text{r}\zeta}{2}\times\nonumber\\&\left(1-2Q\left(\frac{{\min\left(\max\left(a,c_n\right),d_n\right)}-N_\text{r}\zeta}{\sqrt{N_\text{r}\nu^2}}\right)\right)
\nonumber\\&-\frac{\mu_n\sqrt{2N_\text{r}\nu^2}}{\sqrt{\pi}}e^{-\frac{\left({\min\left(\max\left(a,c_n\right),d_n\right)}-N_\text{r}\zeta\right)^2}{2N_\text{r}\nu^2}},
\end{align}
{with $\alpha_n=\frac{e^{\frac{K}{nL}}-1}{\sqrt[1-\vartheta]{\frac{\epsilon P^{\text{cons}}}{(P^{\text{max}})^\vartheta}}}$, $\mu_n=\sqrt[1-\vartheta]{\frac{\epsilon P^{\text{cons}}}{(P^{\text{max}})^\vartheta}}\sqrt{\frac{nL}{2\pi(e^{\frac{2K}{nL}}-1)}},$ $c_n=\alpha_n-\frac{1}{2\mu_n}$ and  $d_n=\alpha_n+\frac{1}{2\mu_n}.$}
\begin{proof}
{See Appendix B.}
\end{proof}

\textbf{\emph{Lemma 4:}} For Rayleigh-fading channels, the probabilities $\mathcal{Y}(a,b,n,L),\forall a,b,n,L,$ are approximated by
\begin{align}\label{eq:eqprobZapproxrayleigh}
&\mathcal{Y}(a,b,n,L)  \simeq  e^{-\frac{\min\left(a,c_n\right)}{\Omega}}\sum_{i=1}^{N_\text{r}}\frac{\left( {\min\left(a,c_n\right)}\right)^{i-1}}{\Omega^{i-1}(i-1)!}
\nonumber\\&-e^{-\frac{\min\left(b,c_n\right)}{\Omega}}\sum_{i=1}^{N_\text{r}}\frac{\left( {\min\left(b,c_n\right)}\right)^{i-1}}{\Omega^{i-1}(i-1)!}
\nonumber\\&
+\left(\frac{1}{2}+\alpha_n\mu_n\right)\times\nonumber\\&\Bigg( e^{-\frac{\min\left(\max\left(a,c_n\right),d_n\right)}{\Omega}}\sum_{i=1}^{N_\text{r}}\frac{\left( {\min\left(\max\left(a,c_n\right),d_n\right)}\right)^{i-1}}{\Omega^{i-1}(i-1)!}\nonumber\\&
-e^{-\frac{\max\left(\min\left(b,d_n\right),c_n\right)}{\Omega}}\sum_{i=1}^{N_\text{r}}\frac{\left( {\max\left(\min\left(b,d_n\right),c_n\right)}\right)^{i-1}}{\Omega^{i-1}(i-1)!}\Bigg)
\nonumber\\&
+\frac{\Omega\mu_n}{(N_\text{r}-1)!}\Bigg(\Gamma\left(N_\text{r}+1,{\min\left(\max\left(a,c_n\right),d_n\right)}\right)\nonumber\\&\,\,\,\,\,\,\,\,\,\,\,\,\,\,\,\,\,\,\,\,\,\,\,\,\,\,\,\,\,\,\,\,\,-\Gamma\left(N_\text{r}+1,{\max\left(\min\left(b,d_n\right),c_n\right)}\right)\Bigg)
\end{align}
\begin{proof}
The proof follows the same approach as in Lemma 3, except that, instead of {the} approximation method of Lemma 1, we use the exact PDF of the sum channel gain as given in (\ref{eq:eqcdfRayleighsum}) and its corresponding CDF $F_G(x)=1-e^{-\frac{1}{\Omega} x}\sum_{i=1}^{N_\text{r}}\frac{ x^{i-1}}{\Omega^{i-1}(i-1)!}$ in Rayleigh-fading conditions.
\end{proof}

From Lemmas 1-4, we can analyze the throughput, the expected delay and the {error} probability of the proposed scheme and compare the system performance with {that of} standard HARQ protocols. For this reason, we consider the {error}-limited expected delay minimization problem
\begin{align}\label{eq:eqproblemfor}
\left\{\begin{matrix}
\mathop {\min }\limits_{q^1,\ldots,q^{M-1}} \,\, \bar\tau \\
 \text{subject}\,\text{to}\,\,\,\,  \Pr\left(\text{error} \right )=\beta \,\,\text{and}\,\, P\le P^\text{max},
\end{matrix}\right.
\end{align}
with {the} quantization boundaries $q^1,\ldots,q^{M-1 }$ being the optimization {variables} and $\beta$ denoting the {error} probability constraint.  Let us define
\begin{align}\label{eq:eqdefpoweralpha}
P_\beta^\text{cons}\doteq \arg\limits_{P^\text{cons}}\left\{\Pr\left(\text{error} \right )=\beta\right\},
\end{align}
i.e., the total consumed power satisfying the {error} probability constraint $\Pr\left(\text{error} \right )=\beta$. Note that the {error} probability (\ref{eq:eqoutageprob1}) is independent of the quantization boundaries $q^m,\forall m$. Hence, using (\ref{eq:eqdefpoweralpha}), $P=\sqrt[1-\vartheta]{\frac{\epsilon P^{\text{cons}}}{(P^{\text{max}})^\vartheta}}$ and the constraints $\Pr\left(\text{error} \right )=\beta \,\,\text{and}\,\, P\le P^\text{max}$, one can determine the required consumed power satisfying the {error} probability constraint as $P^\text{cons}=P_\beta^\text{cons}$ as long as $P_\beta^\text{cons}\le P^\text{max}$. Otherwise, there is no solution for the considered problem formulation (\ref{eq:eqproblemfor}) because the {error} probability constraint can not be satisfied with the considered PA properties. Also, note that (\ref{eq:eqproblemfor}) is a non-convex problem and there is no closed-form solution for the optimal quantization boundaries $q^1,\ldots,q^{M-1}$. However, depending on the maximum number of transmissions $M$, the problem can be effectively solved by exhaustive search {or iterative optimization algorithms, e.g.,} \cite[Algorithm 1]{Tcomkhodemun} {(see Section IV)}. Finally, since the {error} probability is independent of the quantization boundaries, the optimal values of $q^1,\ldots,q^{M-1}$, in terms of expected delay, are the same as those optimized in terms of throughput. In this way, the following lemma shows that the proposed scheme leads to less {error}-limited expected packet transmission delay compared to the standard HARQ protocol where the message is decoded in all transmission rounds and the ACK/NACK signals are fed back in all transmissions except the last round.

\textbf{\emph{Lemma 5:}} The proposed fast HARQ protocol {cannot perform worse than} the standard HARQ protocol, in terms of {error}-limited expected packet transmission delay.

\begin{proof}
For every given values of $q^m,m=1,\ldots,M-1,$ the same {error} probability is achieved by the standard and fast HARQ protocols. Then, setting $q^m=0,\forall m>0$, which is not necessarily optimal in terms of (\ref{eq:eqproblemfor}), fast HARQ is simplified to  standard HARQ. Thus, in the optimal case the proposed fast HARQ protocol outperforms the standard HARQ protocol, in terms of {error}-limited expected delay and, consequently, throughput.
\end{proof}

Finally, defining the relative expected delay gain as
\begin{align}\label{eq:eqrelativedelaygain}
\Xi=\frac{\bar\tau^{\text{standard}}-\bar\tau^{\text{fast}}}{\bar\tau^{\text{{standard}}}},
\end{align}
Lemma 6 quantifies the performance gain of the proposed scheme, compared to the standard HARQ.

\textbf{\emph{Lemma 6.}} Consider moderate/long codewords and a linear decoding delay function $\Lambda(L)=cL$ with $c$ being a constant. At low SNR, the relative expected delay gain of the proposed fast HARQ scheme, compared to the standard HARQ protocol, converges to
\begin{align}\label{eq:Eqlemmafastharq6}
\Xi=\frac{\bar\tau^{\text{standard}}-\bar\tau^{\text{fast}}}{\bar\tau^{\text{standard}}}\mathop  \simeq \limits^{(c)}\frac{c(M-1)}{2+c(M+1)},
\end{align}
independently of the fading model/code rate.
\begin{proof}
At low SNR, all possible transmission rounds are used in both the standard and the fast HARQ protocols. Therefore, with straightforward manipulations in (\ref{eq:eqexpecteddelay}) and using $\Lambda(L)=cL$, the expected {delays for} the fast and standard HARQ schemes are found as $\bar\tau^{\text{fast}}=ML+cML$ and $\bar\tau^{\text{standard}}=ML+\frac{M(M+1)cL}{2}+(M-1)D$, respectively, which leads to the relative expected delay gain as given in (\ref{eq:Eqlemmafastharq6}). {Note that, in }(\ref{eq:Eqlemmafastharq6}), $(c)$ comes from some manipulations and ignoring the feedback delay term in standard HARQ protocol for moderate/long codewords.
\end{proof}
As an example, considering $c=3$, Lemma 6 indicates that with a maximum of $M=2,\ldots,5$ transmissions the fast HARQ scheme results in $27, 42, 52 $ and $60\%$ relative expected delay gain, respectively, which, because the same {error} probability is achieved in both schemes, improves the throughput correspondingly (see Fig. 13). {Also, note that Lemma 6 is consistent with intuition because at low SNRs all message decoding/feedback delay costs are omitted by the fast HARQ protocol. Moreover, the relative performance gain of the fast HARQ, i.e.,} (\ref{eq:Eqlemmafastharq6}), {is independent of the number of information nats per codeword/data rate and increases with the maximum number of transmission rounds. Thus, depending on the code rate and the maximum number of transmission rounds, reasonably low error probability/high throughput may be achieved by fast HARQ at low/moderate SNRs, with considerable improvement in the expected packet delay.}

{\emph{\textbf{Remark 1}}. To have our proposed scheme applicable, the channel should remain constant over $T = (ML+(M-1)D+ \sum_{m=1}^{M-1}\Delta(mL))T_\text{s}$ with $T_\text{s}$ denoting the symbol period. Letting $\Delta(L)=cL$ as in LTE, we have $T = (L (M + \frac{c(M-1)M}{2}) + (M-1)D)T_\text{s}$. Then, as an example, setting $c=1,$ $M=4$ and ignoring $D$, we have $T = 10LT_\text{s}$. If the time slot is 1ms, the channel has to be constant over 10ms. On the other hand, considering systems operating at a carrier frequency of 2.5 GHz and in the case with moving speed of 2 km/h, 45 km/h, and 100 km/h the coherence time is equal to 200 ms, 10 ms, and 4 ms, respectively. Thus, our proposed scheme is well applicable in the cases with stationary or low-speed users. At high speeds, however, either we need to update the estimation of the channel quality/required number of retransmissions after a number of retransmissions (for which the results of} \cite{7556387} {are supportive), or use the initial channel estimate and the temporal correlations between successive channel realizations to predict the required number of retransmissions.}

Note that in our analysis we did not consider the slotted communication setups {motivated by the 802.11 type of communication}. With a slotted communication setup, however, once the feedback is received, the packet is scheduled, but we need to wait for the beginning of the next slot  which affects the packet transmission delay correspondingly. Finally, performance analysis in frequency selective channels, with the data transmitted over a number of parallel channels experiencing independent fading, is an interesting extension of the paper.
\vspace{-0mm}
\subsection{Discussions}
In this part, we discuss the performance of the proposed scheme from different perspectives.
\subsubsection{On the Performance of MIMO-HARQ}
Throughout the paper we concentrated on {SIMO} setups. With a MIMO system, the error probability expression (\ref{eq:errorfiniteblock2}) is rephrased as \cite[Eq. (59)]{6802432}
\begin{align}\label{eq:errorfiniteblockmimo2}
&\delta_\text{quasi-static}(L,R,P)= E\bigg[Q\bigg(\frac{\sqrt{L}\left(C(\textbf{H})-R\right)}{\sqrt{V(\textbf{H})}}\bigg)\bigg],\nonumber\\&
C(\textbf{H})=\log\left|\textbf{I}_{N_\text{r}}+\frac{P}{N_\text{r}}\textbf{H}\textbf{H}^h\right|, \,\nonumber\\&V(\textbf{H})=\min(N_\text{r},N_\text{r})-\sum_{j=1}^{\min(N_\text{r},N_\text{r})}{\frac{1}{(1+\frac{P}{N_\text{r}}\varpi_j)^2}},
\end{align}
with $\varpi_j, j=1,\ldots,\min(N_\text{t},N_\text{r})$, denoting the $\min(N_\text{t},N_\text{r})$ largest eigenvalues of $\textbf{H}\textbf{H}^h$ and $N_\text{t}$ being the number of {transmit} antennas. Thus, one can define  a new set of quantization boundaries on, e.g., $\frac{\sqrt{L}\left(C(\textbf{H})-R\right)}{\sqrt{V(\textbf{H})}},$ and follow the same approach to analyze the system performance in MIMO setups.
\subsubsection{Asymptotic Performance Analysis}
Letting $L\to \infty$ the $Q$-function in (\ref{eq:eqthetaim}) tends towards the step function
\vspace{-0mm}
\begin{align}
&Q\left(\frac{\sqrt{nL}\left(\log\left(1+\sqrt[1-\vartheta]{\frac{\epsilon P^{\text{cons}}}{(P^{\text{max}})^\vartheta}}x\right)-\frac{R}{n}\right)}{\sqrt{1-\frac{1}{\left(1+\sqrt[1-\vartheta]{\frac{\epsilon P^{\text{cons}}}{(P^{\text{max}})^\vartheta}}x\right)^2}}}\right)\nonumber\\&=\left\{\begin{matrix}
1 & \text{if } x\le\frac{e^{\frac{R}{n}}-1}{\sqrt[1-\vartheta]{\frac{\epsilon P^{\text{cons}}}{(P^{\text{max}})^\vartheta}}},\\
0 & \text{otherwise,}
\end{matrix}\right.
\end{align}
with $R\doteq\frac{K}{L}$, and the error probability (\ref{eq:errorfiniteblock2}) maps to the well-known results of
\begin{align}\label{eq:eqlongL}
\delta_\text{quasi-static}(L\to\infty,R,P)=F_G\left(\frac{e^{R}-1}{P}\right),
\end{align}
for the cases with asymptotically long codewords. Also, with straightforward manipulations, (\ref{eq:eqthetaim}) is rephrased as
\begin{align}
\theta_i^m&=F_G\left(\min\left(q^{m-1},\frac{e^{\frac{R}{m}}-1}{\sqrt[1-\vartheta]{\frac{\epsilon P^{\text{cons}}}{(P^{\text{max}})^\vartheta}}}\right)\right)\nonumber\\&-F_G\left(\min\left(q^{m},\frac{e^{\frac{R}{m}}-1}{\sqrt[1-\vartheta]{\frac{\epsilon P^{\text{cons}}}{(P^{\text{max}})^\vartheta}}}\right)\right).
\end{align}
Moreover, using, e.g., Lemma 1 in the cases with long codewords, the {error} probability of the standard and proposed HARQ protocols is approximated by
\begin{align}\label{eq:eqapproxlemmalongL}
\Pr\left(\text{error}\right)\simeq Q\left(\frac{N_\text{r}\zeta-\frac{e^{\frac{R}{M}}-1}{{\sqrt[1-\vartheta]{\frac{\epsilon P^{\text{cons}}}{(P^{\text{max}})^\vartheta}}}}}{\sqrt{N_\text{r}\nu^2}}\right),
\end{align}
with $\zeta$ and $\nu$ given in Lemma 1 for the cases with Rician PDF {(One can also use Lemma 2 to approximate the error probability based on the Gamma-based distribution). It is interesting to note that with asymptotically long codewords and continuous communications model we can follow the same approach as in} \cite[Theorem 1]{Tcomkhodemun} {to show that the proposed fast-HARQ approach leads to the same throughput as in the cases with $\log_2M$-bits quantized CSI feedback, if variable-length coding is used in the INR HARQ.}

\subsubsection{On the Constrained Delay Analysis}
{In practice, the delay used by the upper layer is sometimes measured as the average
delay when the packet is successfully decoded. This metric, which we refer to it as the constrained delay, can be calculated with the same procedure as in} (\ref{eq:eqexpecteddelay}) leading to
\begin{align}\label{eq:eqconstraineddelay}
&\bar\tau_\text{constrained}=\Pr(G\in S^M \&\checkmark \text{in } M) (ML+\Delta(ML))\nonumber\\&+\sum_{m=1}^{M-1}{\Pr(G\in S^m \& \checkmark \text{in } M)(mL+\Delta(mL)}\nonumber\\&{+\sum_{j=m}^M(L+\Delta(jL)+D))}
\nonumber\\&+\sum_{m=1}^{M-1}\sum_{i=m}^{M-1}{\Pr(G\in S^m,\checkmark \text{in } i)(  mL+\Delta(mL)}\nonumber\\&\,\,\,\,\,\,\,\,\,\,\,\,\,\,\,\,\,\,\,\,\,\,\,\,\,\,\,\,\,\,\,\,\,\,\,{+D+\sum_{j=m}^{i-1}{(L+\Delta((j+1)L)+D)}   )}.
\end{align}
{Here, $\Pr(G\in S^m,\checkmark \text{in } i)$ is the probability that the channel is in region $m$ and the message is correctly decoded in round $i$, which can be calculated by following the same analysis as in Section III. Constrained delay analysis and its comparison with the expected delay} (\ref{eq:eqexpecteddelay}) {are presented in Fig. 15. }
\subsubsection{On the Effect of Imperfect CSIR}
{The error probability} (\ref{eq:errorfiniteblock2}) {is a tight approximation for the cases with different levels of CSIR. However, with an imperfect CSIR, the receiver may estimate the instantaneous channel gain in region $m$ to be in region $n\ne m$, which will affect the system performance correspondingly. Following the same procedure as in Section III, the expected delay with imperfect CSIR is given by}
\begin{align}\label{eq:eqestimationerror}
\bar \tau &=\sum_{m=1}^M\sum_{n=1}^M{\Pr(G\in S^m, \tilde G\in S^n)}(mL+\Delta(mL))
\nonumber\\&+\sum_{m=1}^M\sum_{n=1}^{M-1}\sum_{i=n}^{M-1}(L+\Delta((1+i)L))\theta_i^{n,m}
\nonumber\\&+D\sum_{m=1}^M\sum_{n=1}^{M-1}{\Pr(G\in S^m, \tilde G\in S^n)}
\nonumber\\&+
D\sum_{m=1}^M\sum_{n=1}^{M-1}\sum_{i=n+1}^{M-1}\theta_i^{n,m},
\end{align}
where $\tilde G$ denotes the estimated channel. Then,  $\Pr(G\in S^m, \tilde G\in S^n)$ is the probability that the channel $G$ is in region $m$ and estimated to be in region $n.$ Also, $\theta_i^{n,m}$ is the probability that the channel and its estimate are in regions $m$ and $n$, respectively, and the message is not correctly decoded in round $i.$ Considering SISO setups and following the same procedure as in Section III, we have
\begin{align}\label{eq:eqthetaimestimate}
\theta_i^{n,m}&=\int_{q^n}^{q^{n-1}}\int_{q^m}^{q^{m-1}}{f_{G,\tilde G}( x, y)\times} \nonumber\\&{Q\left(\frac{\sqrt{iL}\left(\log\left(1+x\sqrt[1-\vartheta]{\frac{\epsilon P^{\text{cons}}}{(P^{\text{max}})^\vartheta}}\right)-\frac{K}{iL}\right)}{\sqrt{1-\frac{1}{\left(1+x\sqrt[1-\vartheta]{\frac{\epsilon P^{\text{cons}}}{(P^{\text{max}})^\vartheta}}\right)^2}}}\right)\text{d}x\text{d}y},
\end{align}
with $f_{G,\tilde G}(\cdot,\cdot)$ being the joint PDF of $G$ and $\tilde G$ (see Fig. 16 for performance analysis with imperfect CSIR).

{Finally, throughout the paper, we concentrate on the cases with a peak power limit, for which the analytical framework of} \cite{5452208,6802432,6572802} {is applicable. Adaptive power allocation is expected to improve the system performance considerably.}
\section{{Performance Results}}

The expected delay and the throughput of the proposed scheme depend on the characteristics of the message decoding delay function $\Lambda(L)$ in (\ref{eq:eqexpecteddelay}) and (\ref{eq:eqthroughputdef}). Here, we present the results for the cases with a linear decoding delay profile, i.e., $\Lambda(L)=cL, \forall L,$ with $c$ being a constant, which is an appropriate model for different coding schemes. In Figs. {3-5, 7-9, 12-16}, we present the results for the cases with an ideal PA, i.e., we set $\vartheta=0$ and $\epsilon=1$ in (\ref{eq:ampmodeldaniel}). Performance analysis in the cases with an imperfect PA is presented in Figs. {6, 10, 11} where we consider $\vartheta=0.5, \epsilon=0.75$ and $P^\text{max}=48$ dBm, unless otherwise stated. Also, since the variance of the noise is set to 1, we define $P^\text{cons}$ {and $N_\text{r}P^\text{cons}$} (in dB, $\text{SNR}=10\log_{10}P^\text{cons}$ {and $\text{SNR}_\text{total}=10\log_{10}(N_\text{r}P^\text{cons})$}) as the SNR and the total SNR, respectively {(note that with an ideal PA we have $P=P^\text{cons}$ in} (\ref{eq:ampmodeldaniel})). The system performance with a Rayleigh fading channel is studied in Figs. 4b, 5, 16 where we consider $f_{g^i}(x)=\frac{1}{\Omega} e^{-\frac{1}{\Omega} x}$ with $\Omega=1.$ In Figs. 3, 4a, 6-15, however, we consider Rician fading $f_{g^i}(x)=\frac{(k+1)e^{-k}}{\Omega}e^{-\frac{(k+1)x}{\Omega}}I_0\left(2\sqrt{\frac{k(k+1)x}{\Omega}} \right ),\forall i,$ with $k=0.01, \Omega=1$. Figures {6-7} study the system performance in the cases with asymptotically long codewords. Performance analysis of finite blocklength codewords is presented in Figs. {3-5, 8-16}.

In Figs. {4-8}, we consider uniform channel quantization where the quantization boundaries are set such that $\Pr\left(G\in S^m\right)=\frac{1}{M},\forall m.$ {Also, Fig. 16 presents the results for the cases with given quantization boundaries.} In Figs. {9-15}, however, we optimize the quantization boundaries $q^m,\forall m$, in terms of throughput/expected delay.  In our setup, the number of optimization parameters is {small} enough to {allow the} use {of} an exhaustive search, which is what we have used for our simulations. In addition, for faster convergence, we have repeated the simulations by using the iterative algorithm of \cite{Tcomkhodemun} with different initial settings. {In words,} \cite[Algorithm 1]{Tcomkhodemun} {is based on the following procedure. Start the algorithm by considering a number of random possible solutions (in our problem formulation the quantization boundaries $q^i, i=1,\ldots,M$). In each iteration, we determine the best solution, referred to as the queen, that results in the best value of the considered optimization function (in our setup,} (\ref{eq:eqexpecteddelay}){), compared to other considered solutions. Then, we keep the queen for the next iteration and
create a number of solutions around the queen. This is achieved by applying small modifications to the queen.
 Also, to avoid local minima, in each iteration a number of solutions are selected randomly and the iterations continue for a number of times considered by the algorithm designer. Running all considered iterations, the queen is returned as the solution of the optimization problem (see} \cite[Algorithm 1]{Tcomkhodemun} {for more detail).} In all cases, the results {of the exhaustive search and} \cite[Algorithm 1]{Tcomkhodemun} match, with
high accuracy, which is an indication of a reliable result.

{Throughout the paper, we used different approximations. Figure 3 verifies the tightness of these approximations in the cases with Rician PDF, ideal PA, $L=1000$ cu,  $K=500$ nats, $N_\text{r}=50,$ and $M=1, 2.$ Particularly, the figure compares the exact error probability obtained by numerical evaluation of} (\ref{eq:eqprobZ}) {with those obtained by the CLT-, Gamma- and linearization-based approximation schemes of Lemmas 1-3. Also, the figure studies the effect of the third-order term $\frac{\log L}{2L}$ in} (\ref{eq:errorfiniteblock2}) {on the error probability. Then,} in Figs. 4a and 4b, we {study the expected delay and} evaluate the tightness of the approximation schemes of Lemmas 1-4, for the cases with Rician and Rayleigh fading models, respectively. Here, {the results are presented for the cases with} $L=1000$ cu, $K=500$ nats, $N_\text{r}=12, D=40$ cu, $\text{SNR}=0$ dB, $c=0.5$, an ideal PA and different maximum number of transmissions $M$. {Finally,} considering $L=500, 1000$ cu, $K=500$ nats, $D=40$ cu, $\text{SNR}=4$ dB, $M=3,  c=0.5,$ an ideal PA and Rayleigh fading, Fig. {5} presents the expected delay versus the number of {receive} antennas $N_\text{r}$ and evaluates the tightness of the approximation schemes of Lemmas {1, 3, 4}.

\begin{figure}
\vspace{-1mm}
\centering
  \includegraphics[width=0.99\columnwidth]{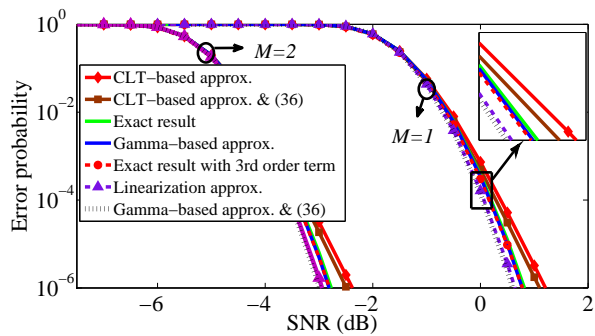}\\\vspace{-1mm}
\caption{On the tightness of the approximation results. Rician PDF, ideal PA, $L=1000$ cu, $K=500$ nats, $N_\text{r}=50,$ $M=1, 2.$}\label{figure111}
\vspace{-1mm}
\end{figure}

\begin{figure}
\vspace{-1mm}
\centering
  \includegraphics[width=0.99\columnwidth]{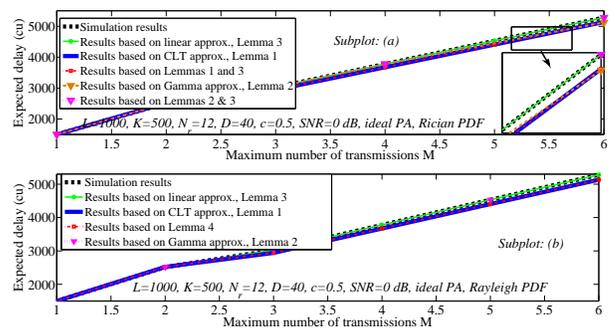}\\\vspace{-1mm}
\caption{Expected delay of the proposed HARQ scheme versus different maximum numbers of transmissions in the cases with (a): Rician and (b): Rayleigh fading distributions. Ideal PA, $L=1000$ cu, $K=500$ nats, $N_\text{r}=12, D=40$ cu, $\text{SNR}=0$ dB and $c=0.5$.}\label{figure111}
\vspace{-1mm}
\end{figure}

\begin{figure}
\vspace{-1mm}
\centering
  \includegraphics[width=0.99\columnwidth]{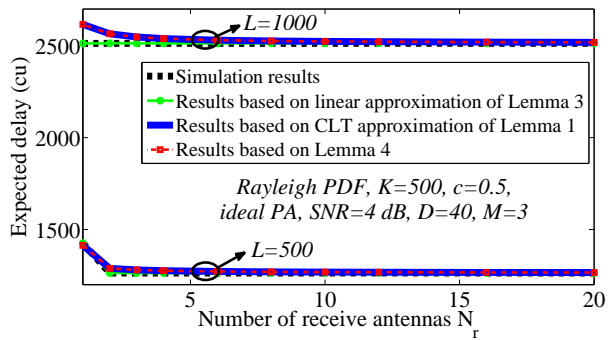}\\\vspace{-1mm}
\caption{Expected delay versus the number of {receive} antennas. Rayleigh fading channel, ideal PA, $L=500, 1000$ cu, $K=500$ nats, $D=40$ cu, $\text{SNR}=4$ dB, $M=3,  c=0.5$.}\label{figure111}
\vspace{-1mm}
\end{figure}

\begin{figure}
\vspace{-1mm}
\centering
  \includegraphics[width=0.99\columnwidth]{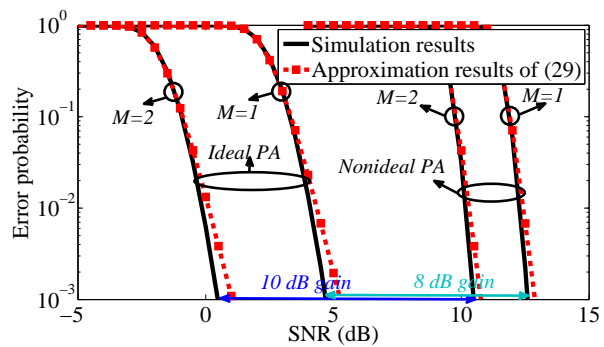}\\\vspace{-1mm}
\caption{{Error} probability versus the $\text{SNR}= 10\log_{10} P^\text{cons}$. Rician fading channel, asymptotically long codeword scenario, $R=1$ npcu, $N_\text{r}=40$, $M=1, 2$.}\label{figure111}
\vspace{-1mm}
\end{figure}

\begin{figure}
\vspace{-1mm}
\centering
  \includegraphics[width=0.99\columnwidth]{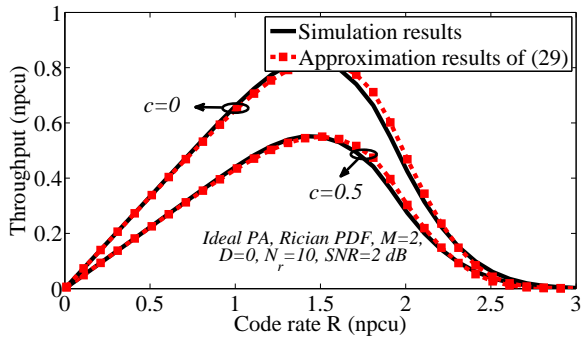}\\\vspace{-1mm}
\caption{Throughput versus the code rate $R$. Rician fading channel, ideal PA,  asymptotically long codeword scenario, $N_\text{r}=10$, $M=2$, $ \text{SNR}=2$ dB, $D=0$.}\label{figure111}
\vspace{-1mm}
\end{figure}

Figures {6-7} study the system {error} probability and throughput, respectively, for the cases with asymptotically long codewords and compare the simulation results with those obtained via {the} approximation {in} (\ref{eq:eqapproxlemmalongL}). Here, the results are presented for the cases with Rician fading and different code rates $R=\frac{K}{L}$.

Considering Rician PDF and an ideal PA, Fig. {8} demonstrates the system throughput versus the codeword length in the cases with $R=1$ npcu, $M=2,$ $D=40$ cu and total SNR $5$ dB. Also, the figure compares the finite blocklength results  with those achieved via Shannon's capacity formula for the cases with asymptotically long codewords. { Then, for given numbers of nats per codeword $K=500, 1000,$ Fig. 9 studies the throughput-expected delay tradeoff of the proposed scheme in the cases with Rician channel, ideal PA, $N\text{t}=6, c = 3, M = 2$ and $D = 40$.}

Considering a Rician channel, non-ideal PA, $D=40$ cu, $N_\text{r}=5, K=L, c=0.5,$ $L=1000$ cu and $M=3$, Fig. {10} demonstrates the optimal quantization boundaries $q^m,\forall m,$ optimizing the expected delay/throughput. Then, Figs. {11a and 11b} respectively study the {error} probability and the throughput of a Rician channel with different PA models/finite blocklength. Here, the results are presented for $K=250$ nats, $L=500$ cu, $D=0, N_\text{r}=3$ and $c=0.5.$

Comparison between the performance of the proposed fast HARQ and standard HARQ is presented in {Figs. {12-13}}, where we{, respectively,} demonstrate the expected delay {and the throughput} for the cases with Rician PDF, an ideal PA, $K=500$ nats, $L=1000$ cu, $D=40$ cu, $N_\text{r}=3, 6$ and $c=3$ as in LTE. Also, the {figures verify} the validity of Lemmas {5-6}.

{With our proposed scheme, there is a small but still nonzero probability of having a packet decoded successfully with a smaller number of retransmission rounds than estimated based on the channel quality. This may lead to unnecessary energy consumption. Considering ideal PA, $c=0.5, L=1000,
K=500, $ and $D=40,$ Fig.  {14a} studies the probability of this unnecessary transmissions in the cases with Rician PDF and $M=2, 3.$ Also, Fig. {14b} shows the unnecessarily consumed energy, {which can be calculated with the same procedure as in Section III,} normalized by the packet length.
Here, the results are presented for the cases with quantization boundaries optimized in terms of} (\ref{eq:eqproblemfor}).


{In Fig. 15, we  compare the expected delay} (\ref{eq:eqexpecteddelay}) {and the constrained delay} (\ref{eq:eqconstraineddelay}) {for Rician channels. Here, the results are presented for the cases with an ideal PA, $c=3,$ $M=2,$ $L=1000,$ $K=1000,$ $N_\text{r}=3$ and $D=40$.}
{Then, considering a SISO setup with Rayleigh fading and the channel model of} \cite{5466522} {with pilot signals of unit power, Fig. 16 uses} (\ref{eq:eqestimationerror})-(\ref{eq:eqthetaimestimate}) {to evaluate the sensitivity of the proposed scheme to CSIR accuracy. Particularly, the figure shows the expected delay as a function of the number of pilot symbols in the cases with $L=1000, K=500, M=2$ and $c=1$.}


According to these figures, the following conclusions can be drawn:

\begin{figure}
\vspace{-1mm}
\centering
  \includegraphics[width=0.99\columnwidth]{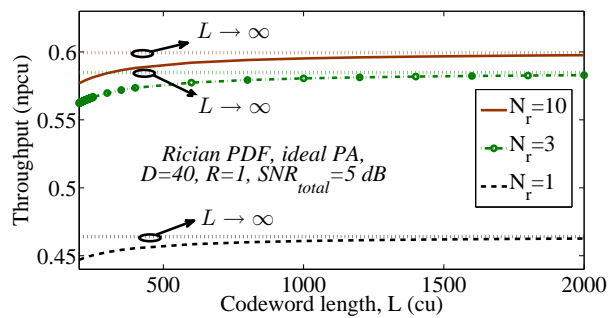}\\\vspace{-1mm}
\caption{Throughput for different codewords lengths. Rician fading channel, ideal PA, $D=40$ cu, $M=2,  c=0$, $R=1$ npcu, $\text{SNR}=5$ dB.}\label{figure111}
\vspace{-1mm}
\end{figure}

\begin{figure}
\vspace{-1mm}
\centering
  \includegraphics[width=0.99\columnwidth]{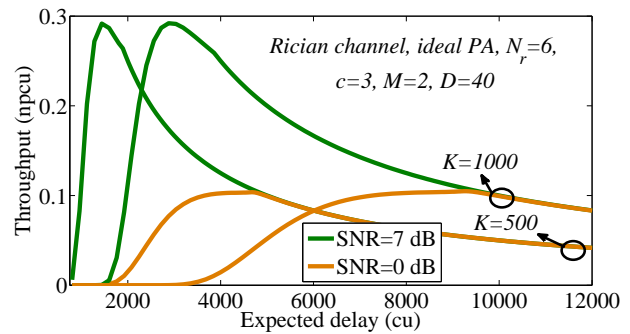}\\\vspace{-1mm}
\caption{On the throughput-delay tradeoff. Rician channel, ideal PA, $N_\text{r}=6, c=3, M=2, D=40.$}\label{figure111}
\vspace{-1mm}
\end{figure}

\begin{figure}
\vspace{-1mm}
\centering
  \includegraphics[width=0.99\columnwidth]{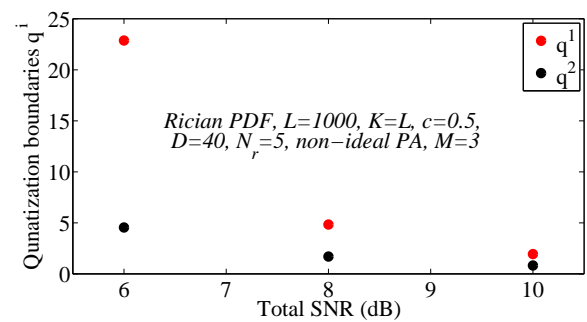}\\\vspace{-1mm}
\caption{Optimal quantization boundaries $q^i$ versus the total SNR $10\log_{10} (N_\text{r}P^\text{cons})$. Rician fading channel, non-ideal PA, $L=1000$ cu, $K=1000$ nats, $D=40$ cu, $M=3,  c=0.5$, $N_\text{r}=5$.}\label{figure111}
\vspace{-1mm}
\end{figure}

\begin{figure}
\vspace{-1mm}
\centering
  \includegraphics[width=0.99\columnwidth]{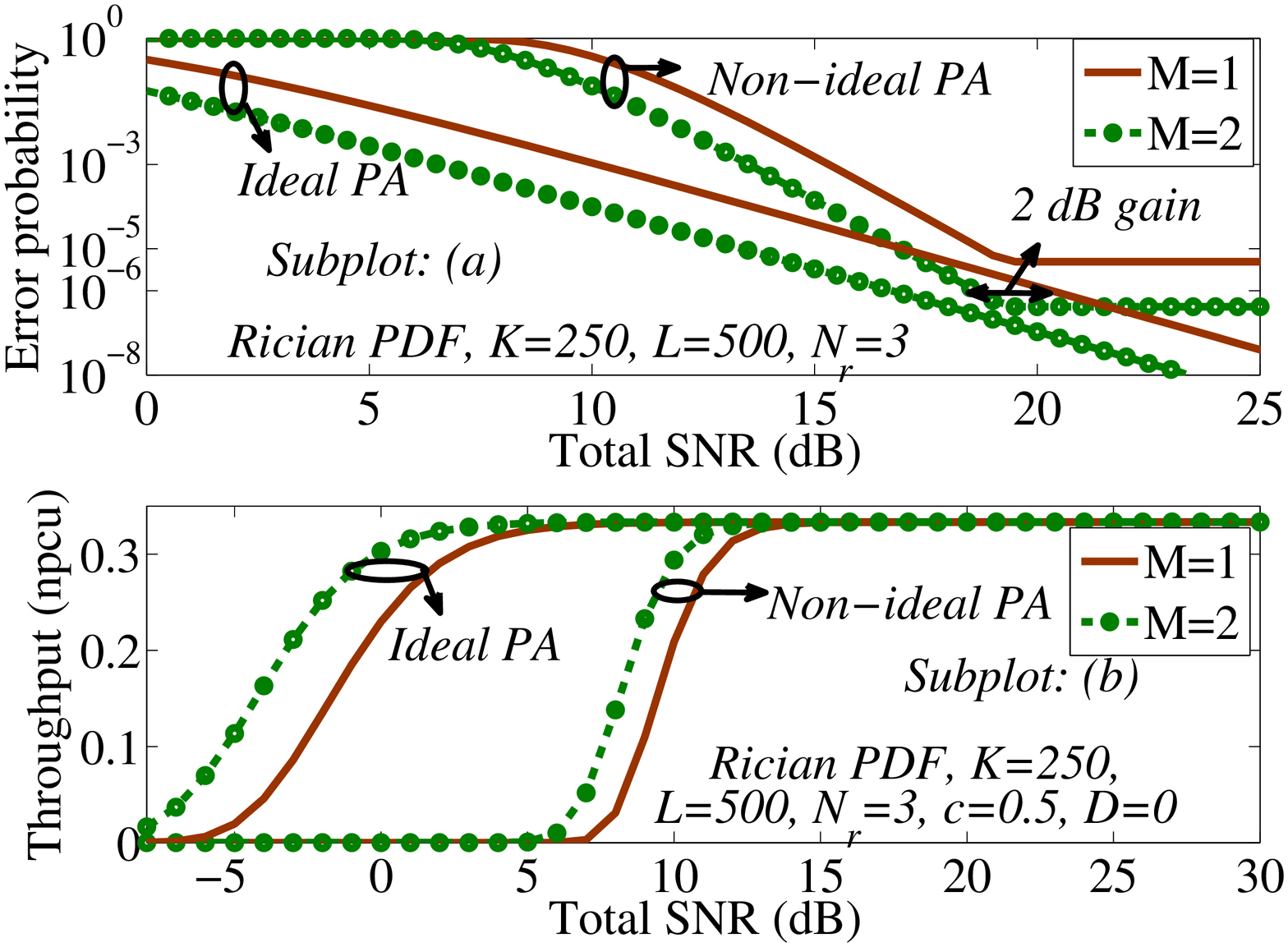}\\\vspace{-1mm}
\caption{On the effect of PA efficiency on the (a): {error} probability and (b): throughput. Rician PDF,  $L=500$ cu, $K=250$ nats, $D=0$ cu, $M=1,2,  c=0.5$, $N_\text{r}=3$.}\label{figure111}
\vspace{-1mm}
\end{figure}


\begin{figure}
\vspace{-1mm}
\centering
  \includegraphics[width=0.99\columnwidth]{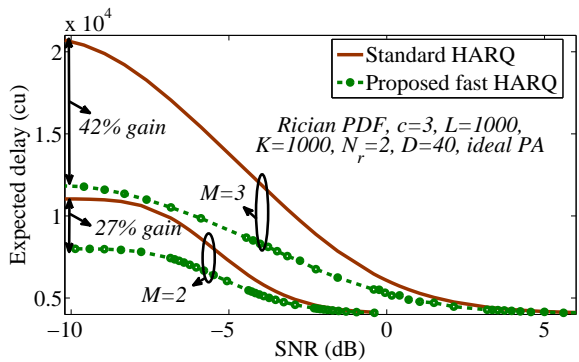}\\\vspace{-1mm}
\caption{Comparison between the delay performance of the proposed and standard HARQ. Rician PDF, ideal PA, $L=1000$ cu, $K=1000$ nats, $M=2,3, c=3, D=40$ cu, $N_\text{r}=3$.}\label{figure111}
\vspace{-1mm}
\end{figure}

\begin{figure}
\vspace{-1mm}
\centering
  \includegraphics[width=0.99\columnwidth]{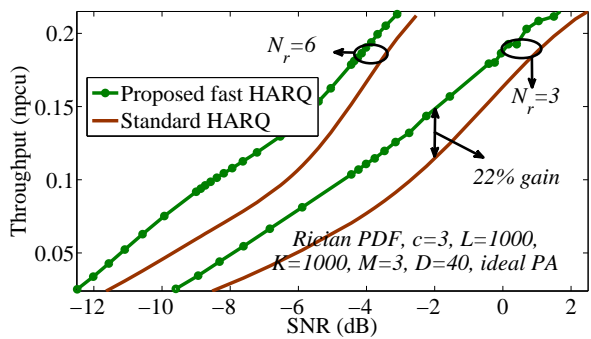}\\\vspace{-1mm}
\caption{Comparison between the end-to-end throughput performance of the proposed and standard HARQ. Rician PDF, ideal PA, $L=1000$ cu, $K=1000$ nats, $M=3, c=3, D=40$ cu, $N_\text{r}=3, 6$.}\label{figure111}
\vspace{-1mm}
\end{figure}

\begin{figure}
\vspace{-1mm}
\centering
  \includegraphics[width=0.99\columnwidth]{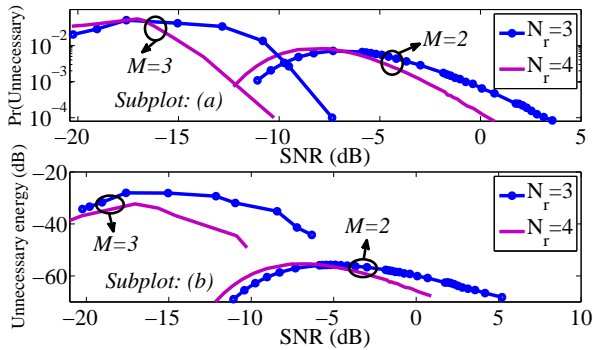}\\\vspace{-1mm}
\caption{{(a): Probability of unnecessary transmission and (b): its corresponding normalized energy in the fast HARQ protocol. Rician PDF, ideal PA, $L=1000$ cu, $K=500$ nats, $D=40$ cu, $M=2,  c=0.5$, $N_\text{r}=3, 4.$}}\label{figure111}
\vspace{-1mm}
\end{figure}

\begin{figure}
\vspace{-1mm}
\centering
  \includegraphics[width=0.99\columnwidth]{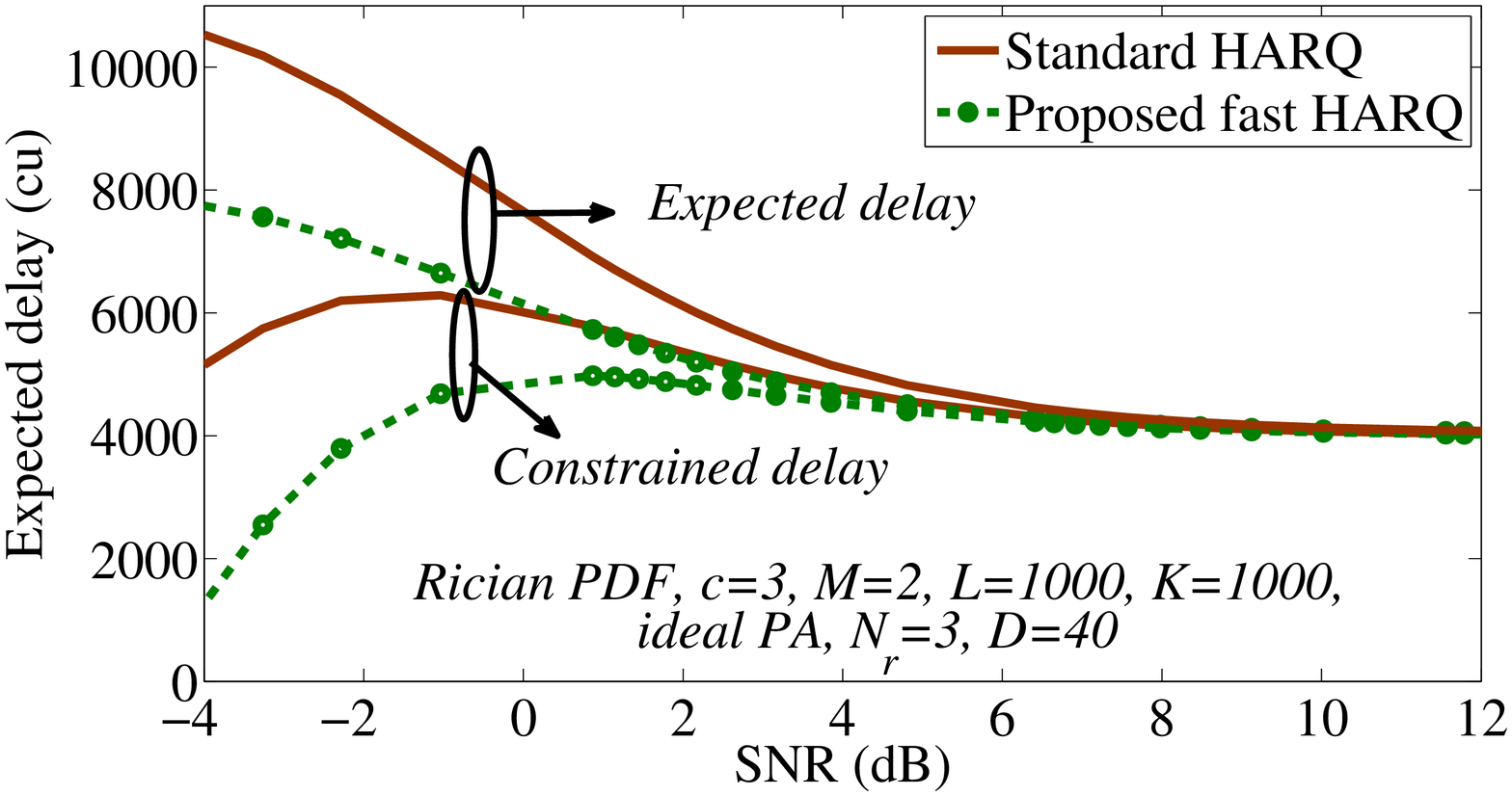}\\\vspace{-1mm}
\caption{Different terms of delay versus the  SNR. Rician PDF, ideal PA, $c=3, M=2, L=1000, K=1000, N_\text{r}=3, D=40.$}\label{figure111}
\vspace{-1mm}
\end{figure}

\begin{figure}
\vspace{-1mm}
\centering
  \includegraphics[width=0.99\columnwidth]{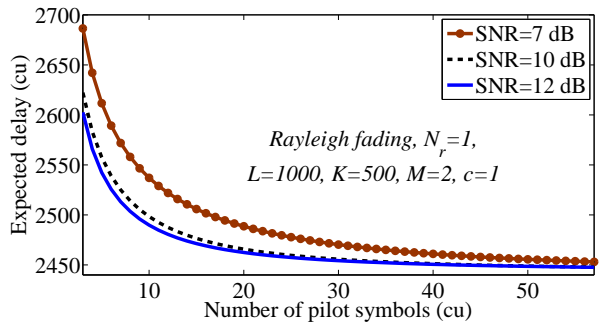}\\\vspace{-1mm}
\caption{Expected delay versus the number of pilot symbols. Rayleigh fading, ideal PA, $L=1000$ cu, $K=500$ nats, $M=2, c=1, q^1=0.25.$}\label{figure111}
\vspace{-1mm}
\end{figure}


\begin{itemize}
  \item As demonstrated in Figs. 3-7, the CLT-, {the Gamma-} and linearization-based approximation approaches of Lemmas 1-4 as well as the approximation scheme of (\ref{eq:eqapproxlemmalongL}) are tight for a broad range of {the} SNR/number of antennas. However, the tightness of {CLT-based} approximations decreases slightly in the cases with high number of transmissions/very low {error} probabilities. Thus, for different parameter settings/fading models, the developed approximations can be effectively applied for the analytical performance evaluation of the proposed and standard HARQ protocols.
  \item For different fading models, the expected packet transmission delay increases (almost) linearly with the maximum number of transmissions $M$ (Figs. {4a-4b}). On the other hand, the {error} probability and the power efficiency are  improved considerably by increasing the maximum number of transmissions. As an example, consider the parameter settings of Fig. 6 and an ideal PA. Then, at {error} probability $10^{-3},$ considering {a} HARQ-based setup with a maximum of $M=2$ transmissions improves the power efficiency by (almost) 4 dB, compared to the cases with open-loop communication ($M=1$). However, because of the quasi-static condition where the channel remains constant in all transmissions, using HARQ does not affect the diversity gain of the system, i.e., the slope of the {error} probability curves in {Figs. 3, 5}. Also, while the expected packet transmission delay decreases with the number of {receive} antennas, it becomes independent of the number of antennas for moderate/large values of $N_\text{r}$ (Fig. {5}). This is intuitive because with {a large} number of antennas the probability of unsuccessful message decoding in the last transmission rounds decreases. Finally, the sensitivity of the expected delay to the number of antennas increases with the codeword length (Fig. {5}).
  \item The inefficiency of the PAs affects the system {error} probability and throughput significantly and should be carefully considered/compensated in the network design. For instance, consider the parameter settings of Fig. 6 and the {error} probability $10^{-3}.$ Then, the PA inefficiency results in $8$ and $10$ dB loss in power efficiency in the cases with $M=1$ and $2$, respectively. Also, with no compensation for the PA {imperfections} the diversity gain of the PA-affected setup is zero as the output power is limited by the PA maximum output power (Fig. {11a}). In the meantime, HARQ can effectively compensate for the PA inefficiency in {error}-constrained scenarios. For example, with the parameter settings of Fig. {11a}, a PA-affected HARQ-based setup with a non-ideal PA requires $2$ dB less SNR, compared to an open-loop system with an ideal PA, to guarantee {a $10^{-6}$} {error} probability. Also, the effect of the PAs inefficiency decreases at high SNRs, as long as the output power is not limited by the PA maximum output power (Figs. {11a-11b}). Intuitively, this is because at high SNRs the effective efficiency of the PAs $\epsilon^\text{effective}=\epsilon\left(\frac{P}{P^\text{max}}\right)^\vartheta$ increases. In the meantime, the PA nonlinearity increases with the SNR, an effect which, while not considered in our work, should be carefully compensated.
  \item As seen in Fig. 7, the throughput increases (almost) linearly with low data rates. This is because with low data rates the message is almost always decoded correctly in the first round(s). On {the} other hand, the {error} probability increases with the data rate and the throughput converges to zero at high data rates. Thus, there is a tradeoff and the maximum throughput is achieved for a finite value of the data rate (Fig. 7). Then, for a given code rate, the throughput is considerably affected by the length of short signals, while its sensitivity to the signals length decreases for signals of moderate/long length (Fig. 8). Particularly, letting $L\to\infty$ the finite blocklength results of \cite{5452208,6802432,6572802} converge to those obtained based on Shannon's results on the achievable rates of long codewords (Fig. 8). Finally, while adding few antennas at the {receiver} increases the throughput {considerably}, compared to a SISO setup, the relative performance gain of adding antennas decreases {for} large $N_\text{r}$ (Fig. 8).
  \item With a given number of information nats per codeword, there is a tradeoff between  {the throughput and the expected delay. With low expected delay, corresponding to low codeword lengths, the data rate is high and the message can not be correctly decoded with high probability. With high codeword length/expected delay, on the other hand, the message is almost always correctly decoded but there are few nats per codeword received by the receiver. Thus, with a given number of nats per codeword, the maximum throughput is achieved with a finite value of the expected delay, and the throughput converges to zero if the expected delay tends to zero or infinity (Fig. 9).}
  \item The optimal quantization boundaries, in terms of expected delay/throughput, converge to zero as the SNR increases (Fig. {10}). This is intuitive because at high SNRs the data is correctly decoded in the first transmission(s) with high probability, and the effect of HARQ decreases. Thus, at high SNR, the relative performance gain of the proposed fast HARQ approach, compared to  standard HARQ, decreases and the quantization boundaries $q^i$ tend towards zero{, which corresponds to} the standard HARQ protocol. This point is also observed in {Figs. {12-13}} where the performance gain of the proposed scheme, compared to standard HARQ, decreases with {the} SNR. However, in harmony with Lemma {5},  fast HARQ always outperforms  standard HARQ, in terms of throughput/expected delay. Also, with low/moderate  power, which is of interest in the large-antenna setups, our proposed scheme leads to {a} remarkable expected delay reduction, compared to  standard HARQ. For instance, considering the parameter settings of Fig. {12}, {a} low SNR regime and a maximum of $M=3$ transmissions, the proposed scheme results in {a} {$42\%$} relative expected delay gain $\Xi$, as also calculated analytically in Lemma {6} (see Lemma {6} and its following discussions). Then, in harmony with Lemma {6}, the performance gain of fast HARQ increases with the maximum number of transmissions and the feedback/message decoding delay. {Also, the relative performance gain is observed at higher SNRs as $M$ increases.} Finally, as it can be seen in Fig. {13}, reasonably high {end-to-end} throughput{ and, consequently, conventional throughput  defined as expected number of successfully received nats per expected packet length, e.g.,} \cite{MIMOARQkhodemun,throughputdef},  are achieved with very low  power, and our fast HARQ scheme outperforms the standard HARQ in terms of end-to-end throughput. {For instance, with the parameter settings of Fig. 13, $N_\text{r}=3$ and SNR$=-2$ dB, the proposed scheme leads to relative throughput gain $\Xi_\text{throughput}=\frac{\eta^\text{fast HARQ}-\eta^\text{standard}}{\eta^\text{fast HARQ}}=22\%.$ Also, it is straightforward to show that, with the parameter settings of Fig. 13, proposed fast HARQ scheme, low/moderate SNRs and different values of $M$, the conventional throughput} \cite{MIMOARQkhodemun,throughputdef} {is $4$ times larger than the end-to-end throughput.}
  \item As seen in Figs. {14a and 14b}, the probability of unnecessary transmission and its corresponding consumed energy are negligible for a broad range of values of the SNR, and the fast HARQ scheme outperforms the standard HARQ, in terms of expected delay (Lemma {5}). {This point leads to an interesting conclusion as follows. As an alternative approach, one may consider a setup where, sending the estimated required retransmission signals together, the receiver tries decoding in all retransmissions. However, as seen in Fig. 14a, the probability of having a packet decoded successfully with a smaller number of retransmission rounds than estimated based on the channel quality is very low. Thus, this alternative scheme will lead to (almost) the same performance as our proposed scheme}.
  \item {The same techniques as in Section III can be well applied for the analysis of other metrics, e.g., constrained delay (Section III.A.3, Fig. 15). Then, at low SNRs, there is low successful message decoding probability and, consequently, constrained delay. At high SNRs, on the other hand, with high probability the message is correctly decoded in the first round and both the expected and the constrained delay converge to $(1+c)L.$ Thus, depending on the channel condition, the maximum constrained delay may be observed at a finite SNR.}
\item {As demonstrated in Fig. 16, for a very few number of pilot symbols, compared to the packet length, the system performance becomes insensitive to imperfect CSIR. Also, the sensitivity of the system performance to the imperfect CSIR decreases with the SNR.}
\end{itemize}

\section{Conclusion}
This paper studied the performance of HARQ-based communication setups in delay-constrained scenarios. Particularly, we developed and analyzed a fast HARQ approach in which, to save on the end-to-end packet transmission delay, different terms of message decoding and feedback delays are reduced.  As demonstrated, our proposed scheme results in a considerable expected delay improvement, compared to standard HARQ protocols. Also, {the throughput and
the error probability are sensitive to the length of short signals while their sensitivity to the signals length decreases for long signals.} Finally, the performance of the HARQ-based setups is considerably affected by the PA's inefficiency, the effect of which should be carefully compensated in the network design.
\appendices
\vspace{-0mm}
\section{Proof of Lemma 1}
Using the {Central Limit} Theorem (CLT) for moderate/large number of antennas, the random variable  $G=\sum_{i=1}^{N_\text{r}}{g^i},$ is approximated by the Gaussian random variable $\mathcal{Z}\sim\mathcal{N}(N_\text{r}\zeta,N_\text{r}\nu^2)${, where,} from (\ref{eq:eqRicianpdf}), $\zeta$ and  $\nu^2$ are, respectively, determined by
\begin{align}\label{eq:eqmeanCLTmmw}
\zeta&=\int_0^\infty{xf_{g^i}(x)\text{d}x}\nonumber\\&=\frac{(k+1)e^{-k}}{\Omega}\int_0^\infty{xe^{-\frac{(k+1)x}{\Omega}}I_0\left(2\sqrt{\frac{k(k+1)x}{\Omega}} \right )\text{d}x},
\end{align}
and
\begin{align}\label{eq:eqmeanCLTmmw}
&\nu^2=\rho-\zeta^2,\,
\nonumber\\&\rho=\int_0^\infty{x^2f_{g^i}(x)\text{d}x}\nonumber\\&=\frac{(k+1)e^{-k}}{\Omega}\int_0^\infty{x^2e^{-\frac{(k+1)x}{\Omega}}I_0\left(2\sqrt{\frac{k(k+1)x}{\Omega}} \right )\text{d}x},
\end{align}
which, using some manipulations and the properties \cite[Eq. (03.02.26.0002.01)]{wolframwebsite}
\begin{align}\label{eq:eqbesselhypppp}
I_n(x)=\frac{1}{\Gamma(n+1)}\left(\frac{x}{2}\right)^n \prescript{}{0}F_{1}\left(n+1;\frac{x^2}{4}\right),
\end{align}
and \cite[Eq. (7.522.5)]{bookhypergeometric}
\begin{align}
&\int_0^\infty{e^{-x}x^{\nu-1}}\prescript{}{s}F_{t}(a_1,\ldots,a_s;b_1,\ldots,b_t;\alpha x)\text{d}x
\nonumber\\&=\Gamma(\nu)\prescript{}{s+1}F_{t}(\nu,a_1,\ldots,a_s;b_1,\ldots,b_t;\alpha),\nonumber
\end{align}
are determined as stated in the lemma.
\vspace{-0mm}
\section{Proof of Lemma 2}
To prove the lemma, we linearize the function $Q\Bigg(\frac{\sqrt{nL}\left(\log\left(1+\sqrt[1-\vartheta]{\frac{\epsilon P^{\text{cons}}}{(P^{\text{max}})^\vartheta}}x\right)-\frac{K}{nL}\right)}{\sqrt{1-\frac{1}{\left(1+\sqrt[1-\vartheta]{\frac{\epsilon P^{\text{cons}}}{(P^{\text{max}})^\vartheta}}x\right)^2}}}\Bigg)$ at point $\alpha_n=\frac{e^{\frac{K}{nL}}-1}{\sqrt[1-\vartheta]{\frac{\epsilon P^{\text{cons}}}{(P^{\text{max}})^\vartheta}}}$ and write
\begin{align}\label{eq:eqlinearQfuncW}
&Q\left(\frac{\sqrt{nL}\left(\log\left(1+\sqrt[1-\vartheta]{\frac{\epsilon P^{\text{cons}}}{(P^{\text{max}})^\vartheta}}x\right)-\frac{K}{nL}\right)}{\sqrt{1-\frac{1}{\left(1+\sqrt[1-\vartheta]{\frac{\epsilon P^{\text{cons}}}{(P^{\text{max}})^\vartheta}}x\right)^2}}}\right)\simeq U_n(x)\nonumber\\& U_n(x)= \left\{\begin{matrix}
1 & x\le c_n, \\
\frac{1}{2}-\mu_n(x-\alpha_n) & x\in\left[c_n,d_n\right],\\
0 & x\ge d_n,
\end{matrix}\right. \nonumber\\&c_n=\alpha_n-\frac{1}{2\mu_n}, d_n=\alpha_n+\frac{1}{2\mu_n},
\end{align}
where
\begin{align}
\mu_n&=-\frac{\partial \left(Q\left(\frac{\sqrt{nL}\left(\log\left(1+\sqrt[1-\vartheta]{\frac{\epsilon P^{\text{cons}}}{(P^{\text{max}})^\vartheta}}x\right)-\frac{K}{nL}\right)}{\sqrt{1-\frac{1}{\left(1+\sqrt[1-\vartheta]{\frac{\epsilon P^{\text{cons}}}{(P^{\text{max}})^\vartheta}}x\right)^2}}}\right)\right)}{\partial x}\bigg|_{x=\alpha_n}\nonumber\\&=\sqrt[1-\vartheta]{\frac{\epsilon P^{\text{cons}}}{(P^{\text{max}})^\vartheta}}\sqrt{\frac{nL}{2\pi(e^{\frac{2K}{nL}}-1)}}\nonumber
\end{align}
is the value of the derivative of $Q\left(\frac{\sqrt{nL}\left(\log\left(1+\sqrt[1-\vartheta]{\frac{\epsilon P^{\text{cons}}}{(P^{\text{max}})^\vartheta}}x\right)-\frac{K}{nL}\right)}{\sqrt{1-\frac{1}{\left(1+\sqrt[1-\vartheta]{\frac{\epsilon P^{\text{cons}}}{(P^{\text{max}})^\vartheta}}x\right)^2}}}\right)$ at point $x=\alpha_n.$ In this way, (\ref{eq:eqprobZ}) is approximated as
\begin{align}\label{eq:eqprobZapprox0}
&\mathcal{Y}(a,b,n,L)\mathop  \simeq \limits^{(a)}\int_a^b{f_G(x)U_n(x)\text{d}x}=\int_{\min\left(a,c_n\right)}^{\min\left(b,c_n\right)}f_G(x)\text{d}x\nonumber\\&+\int_{\min\left(\max\left(a,c_n\right),d_n\right)}^{\max\left(\min\left(b,d_n\right),c_n\right)}f_G(x)\left(\frac{1}{2}+\alpha_n\mu_n- \mu_n x\right)\text{d}x\nonumber\\&
\mathop  \simeq \limits^{(b)}
\int_{\min\left(a,c_n\right)}^{\min\left(b,c_n\right)}f_\mathcal{Z}(x)\text{d}x\nonumber\\&+\int_{\min\left(\max\left(a,c_n\right),d_n\right)}^{\max\left(\min\left(b,d_n\right),c_n
\right)}f_\mathcal{Z}(x)\left(\frac{1}{2}+\alpha_n\mu_n- \mu_n x\right)\text{d}x
\nonumber\\&=\mathcal{U}(a,b,n,L),
\end{align}
{with $\mathcal{U}(a,b,n,L)$ defined in} (\ref{eq:eqprobZapprox}). Here, 
$(a)$ is obtained by (\ref{eq:eqlinearQfuncW}), $(b)$ is based on Lemma 1 and the last equality comes from some manipulations and the CDF of the Gaussian random variable.
\vspace{-0mm}

\bibliographystyle{IEEEtran} 
\bibliography{masterfastHARQ}
\vfill
\end{document}